\documentclass[usenatbib]{mnras}
\usepackage{amsmath}
\usepackage{amssymb} 
\usepackage{graphicx}
\usepackage{caption}
\usepackage{subcaption}
\PassOptionsToPackage{hidelinks}{hyperref}

	\title[Nonresonant interactions with Alfv\'en waves]{A parametric study of population inversions in relativistic plasmas through nonresonant interactions with Alfv\'en waves and their applications to Fast Radio Bursts}

\author[K. Long \& A. Pe'er]{
Killian Long,$^{1}$\thanks{E-mail: killian.long@umail.ucc.ie (KL)}
Asaf Pe'er,$^{2}$
\\
$^{1}$Department of Physics, University College Cork, Cork, Ireland\\
$^{2}$Bar-Ilan University, Ramat-Gan 5290002, Israel
}

\date{Accepted 26 February 2025. Received 08 February 2025; in original form 28 October 2024}

\pubyear{\the\year{}}
\begin{document}
\label{firstpage}
\pagerange{\pageref{firstpage}--\pageref{lastpage}}
\maketitle

	\begin{abstract}
		Synchrotron maser emission is a leading candidate to explain the coherent emission from Fast Radio Bursts (FRBs). This mechanism requires a population inversion in order to operate. We show that nonresonant interactions between Alfv\'{e}n waves and a relativistic plasma result in the formation of population inversions across a wide range of magnetizations, $\sigma\gtrsim10^{-4}$, and temperatures, $10^{-2} \leq k_bT/mc^2 \leq 3$, spanning the parameters expected in FRB environments. We calculate the fraction of energy contained in the inversion across the whole of this parameter space for the first time and we show that energy fractions of $f_{inv}\gtrsim10^{-2}$ are achieved for high magnetizations $\sigma >1$. The population inversion forms on time-scales compatible with the typical dynamical time-scales of magnetars for all magnetizations. Furthermore, we provide physical explanations for the behaviour of the interaction in different magnetization regimes, and identify the important characteristic values at which this behaviour changes. We also show that the mechanism is capable of producing an FRB signal at GHz frequencies in a relativistic magnetar wind close to the light cylinder and that this signal can escape the magnetar environment without significant damping. 
	\end{abstract}
 

 \section{Introduction}

The identity of the emission mechanism responsible for Fast Radio Bursts (FRBs) is an open question, prompting the development of numerous theories \citep[for a summary, see][]{2019PhR...821....1P}. All models share the need to explain the extremely high FRB brightness temperatures of up to $T_b \sim 10^{36}\text{ K}$ \cite[e.g.][]{2016MPLA...3130013K}, far in excess of those achievable by incoherent radiation. One of the primary coherent emission mechanisms invoked to explain the observed FRB signal is synchrotron maser emission (SME) \cite[e.g.][]{Lyubarsky2014, 2018ApJ...864L..12L, Metzger2019}. 
 
 The synchrotron maser, or its mildly relativistic equivalent the cyclotron maser, is a proposed emission mechanism not only for FRBs, but for many other coherent astrophysical phenomena such as Auroral Kilometric Radiation (AKR) in the Earth's magnetosphere \citep{1979ApJ...230..621W, Wu1985}, Jovian Decametric Radiation \cite[e.g.][]{1982AuJPh..35..447H}, and blazars \citep{2005ApJ...625...51B}. 
 SME requires a certain set of plasma conditions: (i) an inverse population of energetic electrons, and (ii) a background magnetic field. Once these conditions are satisfied, the interaction between the inverse electron population and electromagnetic waves in the plasma results in negative absorption and stimulated emission for certain plasma parameters \citep{Wu1985,2006AaARv..13..229T}. 
	
To achieve the necessary population inversion, the particle distribution must either grow faster than $E^2$ \citep{2002ApJ...574..861S}, or satisfy $\frac{\partial F}{\partial v_{\perp}}>0$ \citep{ Wu1985}. Here, $F(v_{\perp},v_{\parallel})$ is the particle distribution function, $v_{\perp} (v_{\parallel})$ is the velocity perpendicular (parallel) to the background magnetic field, and $E$ is the particle's energy. 
The mechanism by which this population inversion forms is crucial to understanding the parameter space where SME is a viable emission mechanism. Various processes have been proposed to pump the electrons into the inverse population. Many of these models, such as the 'loss-cone' \citep{Wu1985, 1976PhFl...19..299F} and 'ring-shell' \citep{1985JGR....90.9650P, 1986ApJ...310..432W} models were developed for non- or mildly relativistic scenarios such as AKR.
	
In the context of FRBs, the environment may be relativistic with temperatures of $k_B T \approx m_e c^2$ \citep[e.g.][]{2020MNRAS.499.2884B, 2020ApJ...896..142B}. Here, $k_B$ is the Boltzmann constant, $m_e$ is the electron mass and $c$ is the speed of light in vacuum. In a previous work \citep{2023PhRvD.107l1301L}, we showed that nonresonant interactions between Alfv\'{e}n waves and a plasma can produce a population inversion capable of supporting SME with a high efficiency, even in such relativistic plasmas. This mechanism does not require a shock wave as in the other primary models for relativistic SME, where a soliton-like structure is formed at the front of a strongly magnetized shock in which particles gyrate around the enhanced magnetic field and form a semi-coherent ring in momentum space \cite[e.g.][]{1988PhFl...31..839A, 1992ApJ...391...73G, 2006ApJ...653..325A, Plotnikov2019}. Similar ring distributions may also be formed through radiation reaction cooling \citep{2023PhRvL.130p5101B, 2024PhPl...31e2112B,2024arXiv240918955B}.
		
Nonresonant interactions between the Alfv\'{e}n waves and the relativistic plasma induce pitch-angle diffusion in low $\beta$ plasmas, where $\beta$ is the ratio of the plasma pressure to the magnetic pressure. In the emission regions of FRBs the environment is highly magnetized, satisfying this condition even at relativistic temperatures. The resulting deformation of the initial particle distribution into a crescent shape in momentum space is capable of supporting a population inversion \citep{Zhao2013, Wu2014}. 

In this work, we build upon the proof of concept results presented in \cite{2023PhRvD.107l1301L} where the full relativistic interaction was studied for the first time. We explore the details of the fraction of the distribution's energy contained in the population inversion as well as the time-scales required. We examine a large parameter space in both temperature $T$ and magnetization $\sigma=\Omega^2/\omega_p^2$ for the first time. Here $\Omega = eB_0/mc$ is the gyration frequency of the plasma as a whole and $\omega_{p} =\sqrt{4\pi n e^2/m}$ is the plasma frequency,  $e$ is the electron charge, $B_0$ is the background magnetic field, and $n$ is the particle number density. The entire magnetization parameter space $\sigma > 10^{-4}$ is investigated at temperatures in the range $10^{-2}\leq\theta \leq3$, where $\theta =\frac{k_BT}{mc^2}$ is the normalized temperature, providing a full picture of the interaction across all the relevant physical parameters.

As in \cite{2023PhRvD.107l1301L}, to examine the problem we consider a relativistic plasma of density $n$ embedded in a background magnetic field $\mathbf{B_0}$. The initial particle distribution $F_0(p_\parallel,p_\perp,\theta_0)$ is taken to be a Maxwell-J\"{u}ttner distribution with temperature $\theta_0$, where $p_\perp = \gamma m v_\perp$ and $p_{||}=\gamma m v_\parallel$ are the relativistic momenta in the perpendicular and parallel directions to the background magnetic field respectively. Here, $\gamma = (1+(p_\perp/mc)^2+ (p_\parallel/mc)^2)^{1/2} $ is the Lorentz factor. The nonresonant interaction between the particles and the waves causes the distribution to evolve, with the details of the development depending on the magnetization and temperature. This evolved distribution is specified at time $t$ by $F(p_\parallel,p_\perp,t)$.  In order for nonresonant interactions to occur, it is necessary for Alfv\'{e}n waves to propagate through the plasma in addition to the steady background magnetic field. In the typical FRB scenario, the Alfv\'{e}n waves are generated by a central neutron star/magnetar, resulting in a wavevector that is expected to be in the parallel direction only. The spectral magnetic and electric fields of the waves are denoted as $B_k$ and $E_k$ respectively, where $k$ is the wavenumber. The waves are assumed to vary slowly in time, and to have a broad spectrum \citep{Wu2007,Yoon2009}.

This paper is organized as follows. In Section \ref{sec:nonresonant} we detail the physical setup of our model and the equations which describe it. The behaviour of the full relativistic quasilinear kinetic equation for different parameters and the numerical methods used to solve the equation are examined in Section \ref{sec:Methods}, with some semi-analytical results for the behaviour of individual terms derived in Appendix \ref{sec:termcomp}. In Section \ref{sec:num} we present the results for both the energy fraction contained in the population inversion and the relevant formation time-scales. These results are discussed and compared to the physical conditions necessary to produce FRBs in Section \ref{sec:discussion}, and the results of the paper are summarized in the conclusion in Section \ref{sec:conclusion}.

\section{Non-resonant interaction of particles with Alfv\'en waves: evolution of the particle distribution function}
\label{sec:nonresonant}
We use kinetic theory to study the interaction between the waves and particles. The quasilinear approximation of this theory \citep{vedenov1961nonlinear, drummond1962anomalous} is used to determine how the particle distribution changes due to the interaction with the Alfv\'{e}n waves. Here the variables of the Vlasov equation are split into slowly varying average and first order fluctuation terms. Ref. \cite{stix1992waves} starts from the Vlasov equation, and provides a complete derivation of the full relativistic quasilinear kinetic equation describing the temporal evolution of the particle's distribution function due to interaction with the Alfv\'en waves at resonance. We added nonresonant terms following \cite{Yoon2009}. The complete equation, that contains both the resonant (line 2) and the nonresonant (line 3) terms and is correct in both relativistic and non-relativistic regimes, is

\begin{eqnarray}
\frac{\partial F}{\partial t}&=&\frac{e^2}{4}\sum_{l=\pm 1}\int d\textbf{k}\frac{1}{p_{\perp}}\left[\left(1-\frac{k_\parallel p_\parallel}{\gamma m\omega}\right)\frac{\partial}{\partial p_\perp}+\frac{k_\parallel p_\perp}{\gamma m\omega}\frac{\partial}{\partial p_\parallel}\right]\nonumber\\&&\times\Bigg\lbrace p_\perp\Bigg[\pi\delta\left(\omega-l\omega_c-\frac{k_\parallel p_\parallel}{\gamma m}\right)|E_k|^2\nonumber\\&&-\frac{\partial}{2\partial\omega}\left(PV\left(\frac{1}{\omega-l\omega_c-\frac{k_\parallel p_\parallel}{\gamma m}}\right)\right)\frac{\partial\left|E_k\right|^2}{\partial t}\Bigg]\nonumber\\&&\times\left[\left(1-\frac{k_\parallel p_\parallel}{\gamma m\omega}\right)\frac{\partial}{\partial p_\perp}+\frac{k_\parallel p_\perp}{\gamma m\omega}\frac{\partial}{\partial p_\parallel}\right]F\Bigg\rbrace.
\label{eq:2}
\end{eqnarray} 
Here $\omega_c = eB_0/\gamma mc$ is the particle gyration frequency; $\omega$ is the Alfv\'{e}n wave frequency (whose dispersion relation is described in Section \ref{sec:Alfven} below); $PV$ denotes a Principal Value; $k_\parallel$ is the wavenumber in the direction parallel to the background magnetic field, and $l$ is the harmonic number. The integration is over the Alfv\'enic wavevector $\textbf{k}$, which is assumed to be in the direction parallel to $\bf{B}_0$, as described above.

The equation is governed by whether the resonance condition $\omega -l\omega_c - \frac{k_\parallel p_\parallel}{\gamma m}=0$ is satisfied. While this is the case in many plasmas, it is not in FRB emission regions. Due to strong magnetic fields, plasmas in these environments have a low $\beta$.  For Alfv\'{e}n waves in such a plasma, the inequalities $\omega_c \gg \omega$ and  $\omega_c \gg \frac{k_\parallel p_\parallel}{\gamma m}$ hold \citep{Wu2007}, resulting in no contribution from the resonant term. The presence of the $\frac{\partial |E_k|^2}{\partial t}$ term requires that the Alfv\'{e}n wave electric field varies temporally in the nonresonant case, a restriction that is not found in the resonant regime. 

 To ensure that the above conditions are always satisfied, the maximum wavenumber of the Alfv\'{e}n wave, $k_{\max}$ must not be too large. Assuming an Alfv\'{e}n relation of $w = kc$ in the highly magnetized and relativistic case, the condition for the nonresonant interaction to remain dominant becomes $\omega_c\gg k_{\max}c$. We can further assume that the Alfv\'{e}n wave is generated by some instability of the neutron star such as a quake \citep{1989ApJ...343..839B}. Thus the wavelength $\lambda = \xi R_*$ will be some fraction $\xi$ of the neutron star radius $R_*$, with a corresponding wavenumber of $k\sim2\pi/(\xi R_*)$. The condition of validity thus becomes $2\pi c/(\omega_cR_*) \ll \xi$. Using fiducial values of $R_* = 10^6\text{ cm}$ and $\omega_c = 10^9 \text{ rad s}^{-1}$, which is appropriate for FRBs, we obtain the condition $\xi \gg 2 \times 10^{-4}\omega_{c,9}^{-1}R_{*,6}^{-1}$, where we use the convention $Q = 10^x Q_x$ in cgs units.  Typical wavelengths of waves in neutron star crusts are estimated to be of the order 
 of $\lambda \sim10^4\text{ cm}$ to $\lambda\sim10^7\text{ cm}$, the equivalent of $\xi \sim 10^{-2} R_{*,6}^{-1}$ to $\xi \sim 10 R_{*,6}^{-1}$  \citep{1989ApJ...343..839B, 2023ApJ...959...34B}. Therefore, the maximum value of $k_{\max}$ will not exceed a fraction of $\omega_c/c$, confirming that the interaction will remain in the nonresonant regime. In low magnetization cases, the value of $k_{\max}$ will be lower due to the reduced Alfv\'{e}n velocity, resulting in an even less restrictive limit.

The nonresonant term can itself be simplified through the approximation \citep{Yoon2009}:
\begin{equation}    
\frac{\partial}{2\partial\omega}\left(PV\left(\frac{1}{\omega-l\omega_c-\frac{k_\parallel p_\parallel}{\gamma m}}\right)\right)\frac{\partial\left|E_k\right|^2}{\partial t}\approx-\frac{1}{2\omega_c^2}\frac{\partial\left|E_k\right|^2}{\partial t}.
\end{equation}

In order to present equation \eqref{eq:2} more clearly, we expand the equation and separate the terms, using the $l=\pm1$ harmonics. The resulting expression is
\begin{align}
\frac{\partial F}{\partial t}&=\frac{7.7\times10^{-5}}{\omega_{c,9}^2}\Bigg\lbrace \left(\frac{I_1\left(2+\frac{q_\perp^2}{\gamma^2}\right)}{  \gamma}-2\frac{I_2q_\parallel}{ \gamma^2}\right)\frac{\partial F}{\partial q_\parallel}\nonumber \\
&+\Bigg[\frac{1}{q_{\perp}}\left(I_3\left(1+2\frac{q_\perp^2}{\gamma^2}\right)+\frac{I_2q_\parallel^2}{\gamma^2 }-\frac{I_1 q_{\parallel} \left(2+\frac{q_\perp^2}{\gamma^2}\right)}{\gamma }\right)\nonumber\\
&-\frac{I_2q_\perp}{\gamma^2}\Bigg]\frac{\partial F}{\partial q_\perp}\nonumber\\
&+\frac{I_2 q_\perp^2}{\gamma^2 }\frac{\partial^2 F}{\partial q_\parallel^2}\nonumber\\
&+\left(I_3+\frac{I_2q_\parallel^2}{\gamma^2}-2\frac{I_1 q_\parallel}{\gamma }\right)\frac{\partial^2 F}{\partial q_\perp^2}\nonumber\\
&+\left(2\frac{ I_1 q_\perp}{\gamma }-2\frac{I_2 q_\perp q_\parallel}{\gamma^2 }\right)\frac{\partial^2 F}{\partial q_\parallel \partial q_\perp}\Bigg\rbrace,
\label{eq:4}
\end{align}  
where the factor $\frac{7.7\times 10^{-5}}{\omega_{c,9}^2} = \frac{e^2}{4 c^2 m^2 \omega_c^2}$, $q=p/mc=\gamma \frac{v}{c}=\gamma\beta$, and $I_1 = \int d\textbf{k}\frac{\partial\left|E_k\right|^2}{\partial t}\frac{ck_\parallel}{\omega}$, $I_2 = \int d\textbf{k}\frac{\partial\left|E_k\right|^2}{\partial t}\frac{c^2k_\parallel^2}{\omega^2}$ and $I_3 = \int d\textbf{k}\frac{\partial\left|E_k\right|^2}{\partial t}$. Equation \eqref{eq:4} is correct in the limit of strong magnetic fields, in both the relativistic and non-relativistic regimes. \footnote{There is a typographical error in equation (3) of \cite{2023PhRvD.107l1301L}, causing the discrepancy with equation \eqref{eq:4} of this work which has extra terms on lines 1 and 2. The equation presented above is the correct version.} 

Each line in equation \eqref{eq:4} describes a different component of the overall evolution, which we respectively denote as $\left(\frac{\partial F}{\partial t}\right)_x$. From top to bottom, these terms describe parallel advection $\left(\frac{\partial F}{\partial t}\right)_{||,1}$, perpendicular advection $\left(\frac{\partial F}{\partial t}\right)_{\perp,1}$, parallel diffusion $\left(\frac{\partial F}{\partial t}\right)_{||,2}$, perpendicular diffusion $\left(\frac{\partial F}{\partial t}\right)_{\perp,2}$ and a mixed term $\left(\frac{\partial F}{\partial t}\right)_{mix}$.

\subsection{Dispersion relation of the Alfv\'en waves}
\label{sec:Alfven}

In order to solve the time evolution of the distribution function described by equation \eqref{eq:4}, one needs to evaluate the integrals $I_1$, $I_2$ and $I_3$. 
The integrals $I_1$ and $I_2$ are calculated using either the non-relativistic Alfv\'{e}n dispersion relation $\omega =kv_A$, where $v_A$ is the Alfv\'{e}n velocity, or the full relativistic solution given by \citep{Asenjo2009,Munoz},
\begin{equation}
\omega^2-c^2k^2=\underset{s}{\sum}\omega_{p(s)}^2\frac{\omega'}{f_s\bar{\gamma}_s \omega'-\Omega_{s}} \label{eq:alf}.
\end{equation}
Here, the subscript $s$ denotes the particle species; $\omega'=\omega - kv_0$; $v_0$ is the particles' bulk velocity (in case it is non-zero); $\omega_{p(s)} =\sqrt{4\pi n_s e^2/m_s}$ is the plasma frequency; $f_s=K_3\left(\frac{m_sc^2}{k_B  T_s} \right)\Big/K_2\left(\frac{m_s c^2}{k_B T_s}\right)$ is the plasma's enthalpy; $\bar{\gamma}_s$ is the species Lorentz factor and $\Omega_s = q_s B_0 / m_s c$ is the species gyration frequency where $q_s$ is the species charge. $K_l$ denotes a modified Bessel function of the second kind of order $l$. 

In the non-relativistic regime the ratio between the integrals is simply given by $I_1 = \frac{v_A}{c}I_2 =\frac{c}{v_A}I_3$. In this limit therefore $I_2>I_1>I_3$ for all $v_A<c$. This relationship also holds for the relativistic case provided that $k_{\max}$ is sufficiently small that the $\omega-k$ relationship remains linear, which is the case for the parameter space examined in this work. 

To demonstrate this, expanding equation \eqref{eq:alf} for a two species plasma gives
\begin{equation}
\omega^2-c^2k^2=\omega_{p-}^2\frac{\omega'}{f_-\bar{\gamma}_-w'-\Omega_{-}}+\omega_{p+}^2\frac{\omega'}{f_+\bar{\gamma}_+w'-\Omega_{+}}, 
\end{equation}
where the subscript $+$ denotes the positively charged species and the subscript $-$ denotes the electrons. Defining $\mu=m_+/m_-$ as the mass ratio of the two species and assuming both species have equal $f$ and $\bar{\gamma}$, the above equation can be written as
\begin{equation}
\omega^2-\omega\left(\frac{\omega_{p+}^2}{f\bar{\gamma}\omega-\Omega_{+}}+\frac{\mu\omega_{p+}^2}{f\bar{\gamma}\omega+\mu\Omega_{+}}\right)-c^2k^2=0, \label{eq:alfstep}
\end{equation}
for particles with zero bulk velocity. In the low frequency limit, $f\bar{\gamma}\omega<\Omega_{+}$, one can make the following approximations:

\begin{eqnarray}
\frac{\omega_{p+}^2}{f\bar{\gamma}\omega-\Omega_{+}}&\approx&-\frac{\omega_{p+}^2}{\Omega_{+}}\left(1+\frac{f\bar{\gamma}\omega}{\Omega_{+}}\right) \nonumber \\
\frac{\mu\omega_{p+}^2}{f\bar{\gamma}\omega+\mu\Omega_{+}}&\approx&\frac{\omega_{p+}^2}{\Omega_{+}}\left(1-\frac{f\bar{\gamma}\omega}{\mu\Omega_{+}}\right).
\end{eqnarray}
The term in brackets in equation \eqref{eq:alfstep} can therefore be written as
\begin{equation}
\left(\frac{\omega_{p+}^2}{f\bar{\gamma}\omega-\Omega_{+}}+\frac{\mu\omega_{p+}^2}{f\bar{\gamma}\omega+\mu\Omega_{+}}\right)\approx -\frac{f\bar{\gamma}\omega \omega_{p+}^2}{\Omega_{+}^2}\left(1+\frac{1}{\mu}\right),
\label{eq:mu}
\end{equation}
and the entire dispersion relation can therefore be approximated as
\begin{equation}
\omega^2\left(1+\frac{ f \bar{\gamma} \omega_{p+}^2}{\Omega_{+}^2}\left(1+\frac{1}{\mu}\right)\right)\approx c^2k^2,
\end{equation}
showing that the relativistic Alfv\'{e}n wave dispersion relation is linear in the low frequency regime of interest to this work, with a corresponding relativistic Alfv\'en velocity of $v_{A,rel} = c\left(1+\frac{ f \bar{\gamma} \omega_{p+}^2}{\Omega_{+}^2}\left(1+\frac{1}{\mu}\right)\right)^{-1/2}$.

\subsection{Spectrum of the Alfv\'{e}n wave}
\label{sec:spectrum}
The outcome of the wave-particle interaction is insensitive to the precise spectrum chosen provided that the Alfv\'en wave dispersion relation is linear, as in this case $I_1 = \frac{v_{A,rel}}{c}I_2 =\frac{c}{v_{A,rel}}I_3$ always holds. We demonstrate this below. Consider waves having a power law spectrum where the spectral electric field is given by $E_k(t) = E_0(t) k^{-\varphi}$. Using $\frac{\partial | E_k(t) |^2}{\partial t} = 2 \Gamma_k E_k(t)^2$, where $\Gamma_k$ is the spectral growth rate of the Alfv\'{e}n wave, the terms $I_1$, $I_2$ and $I_3$ become
\begin{eqnarray}
I_1 &=&2\int d\textbf{k}\Gamma_k E_0^2k_{||}^{-2\varphi}\frac{k_\parallel}{\omega} = \frac{2}{v_{A,rel}}\int d\textbf{k}\Gamma_k E_0^2k_{||}^{-2\varphi}, \nonumber \\ 
I_2 &= &2  \int d\textbf{k}\Gamma_k E_0^2k_{||}^{-2\varphi}\frac{ck_\parallel^2}{\omega^2} = \frac{2c}{v_{A,rel}^2}\int d\textbf{k}\Gamma_k E_0^2k_{||}^{-2\varphi}, \nonumber \\
I_3 & =& 2 \int d\textbf{k}\Gamma_k E_0^2\frac{k_{||}^{-2\varphi}}{c}= \frac{2}{c}\int d\textbf{k}\Gamma_k E_0^2k_{||}^{-2\varphi},	\label{eq:spectrum}
\end{eqnarray}
with ratios that are independent of the power law exponent. Note that we explicitly used the assumption that the Alfv\'en waves propagation direction is  parallel to the background magnetic field. 

The magnitude of each $I_x$ term ($x=1,2,3$), and thus the time-scale and turbulence dependence of the process, is also independent of the precise spectrum provided that: (i) the spectral growth rate $\Gamma_k$ is a constant and (ii) the total energy contained in the spectra in question is the same.

To illustrate a case which satisfies the above requirements, we continue the example of a power law spectrum presented in equation \eqref{eq:spectrum} with the added constraint that $\Gamma_k$ is constant. The $I_x$ terms can be written as
\begin{eqnarray}
I_1 &=&\frac{2\Gamma_kE_0^2}{(-2\varphi +1)v_{A,rel}}\left(k_{\max}^{-2\varphi + 1}-k_{\min}^{-2\varphi + 1}\right), \nonumber\\ 
I_2 &= &\frac{2\Gamma_k c E_0^2}{(-2\varphi +1)v_{A,rel}^2}\left(k_{\max}^{-2\varphi + 1}-k_{\min}^{-2\varphi + 1}\right), \nonumber\\
I_3 & =& \frac{2\Gamma_kE_0^2}{(-2\varphi +1)c}\left(k_{\max}^{-2\varphi + 1}-k_{\min}^{-2\varphi + 1}\right).	
\end{eqnarray}
Comparing this to a flat spectrum ($\varphi = 0$), ratio $I_{x,\text{flat}}/I_{x,\text{power law}}$ given by
\begin{equation}
\frac{I_{x,\text{flat}}}{I_{x,\text{power law}}} = \frac{\left(k_{\max}^{-2\varphi + 1}-k_{\min}^{-2\varphi + 1}\right)}{(k_{\max}-k_{\min})(1-2\varphi)}.
\end{equation}
This is the same factor obtained when comparing the energy density of a flat spectrum $E_k = E_0$ and power law spectrum $E_k = E_0' k^{-\varphi}$:
\begin{eqnarray}
\int dk E_0^2& =& \int dk E_0^{'2}k^{-2\varphi}, \nonumber\\
E_0^2 (k_{\max}-k_{\min}) &=& E_0^{'2}\frac{\left(k_{\max}^{-2\varphi + 1}-k_{\min}^{-2\varphi + 1}\right)}{(1-2\varphi)},\nonumber \\
\frac{E_0^2}{E_0^{'2}} &=& \frac{\left(k_{\max}^{-2\varphi + 1}-k_{\min}^{-2\varphi + 1}\right)}{(k_{\max}-k_{\min})(1-2\varphi)},\label{eq:powerlaw}
\end{eqnarray} 
where we have assumed an equal wavenumber range in both cases. This demonstrates that the magnitudes of the $I_x$ terms, and thus the time-scale of the interaction, are independent of the Alfv\'en wave spectrum in this case.

\section{Methods and analysis}
\label{sec:Methods}
\subsection{Solving the nonresonant wave-particle interaction}
\label{sec:Methods1}
To solve equation \eqref{eq:4}, a number of numerical schemes were used. These include the Crank-Nicolson, upwind differencing and fully implicit finite difference methods \citep{Press2007}. The equation was solved with MATLAB on a term by term basis. The initial conditions are the initial distribution function $F_0$, which is taken to be a Maxwell-J\"{u}ttner distribution with a normalized temperature $\theta =\frac{k_BT}{mc^2}$; and the magnetization $\sigma = \Omega^2/\omega_p^2$. The Alfv\'en wave turbulence is described by the quantity $\eta = \frac{\int d\textbf{k}B_k^2}{B_0^2}$, the ratio of the energy density in the waves to the background field. This ratio is assumed to take the form $\eta(t) = \eta_0 e^{2\Gamma t}$, where $\eta_0$ is the initial turbulence level. Based on the discussion in Section \ref{sec:spectrum}, we take $\Gamma = \Gamma_k$ as a constant for simplicity. To ensure the quasilinear approximation remains valid, we require $\eta<<1$ and $\Gamma << \Omega$. We further assume a single temperature and density describing all species. The equation is evolved in units of $\tau =  \frac{1}{4}\frac{v_{A,rel}^2}{ c^2}\left \langle\frac{\partial \eta}{\partial t}\right \rangle t$, where $\left \langle\frac{\partial \eta}{\partial t}\right \rangle$ is the time averaged value. The state of the distribution depends on the turbulence level of the Alfv\'en waves, and the time to reach a given state depends on the growth rate of this turbulence. The results in Section \ref{sec:energy} are therefore presented in terms of $\eta$ for clarity. Improvements to the stability of the numerical scheme allowed higher resolutions and longer times to be achieved compared to the results presented in our previous proof of concept work \citep{2023PhRvD.107l1301L}.

 The distribution function at a given value of $\tau$ is calculated by solving equation \eqref{eq:4}. This requires calculating the values for the integrals $I_1$, $I_2$ and $I_3$. We use the dispersion relation (equation \eqref{eq:alf}) as discussed in Section \ref{sec:Alfven}. While the values of these integrals determine the time scale for reaching a steady state, their ratio determines the amount of energy available in the population inversion. 

\begin{figure}
	\centering
	\includegraphics[width=\columnwidth]{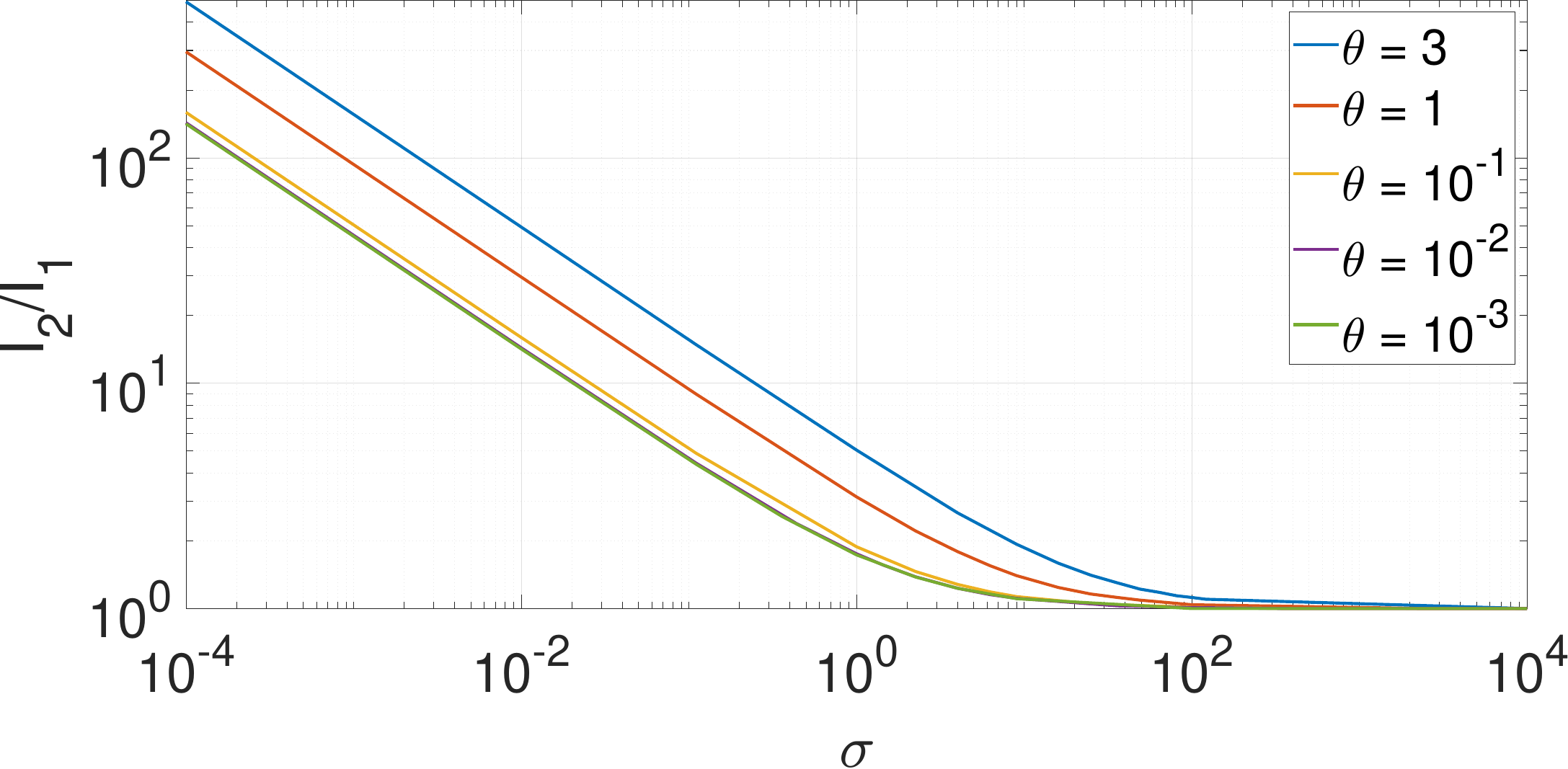}
	\caption{The value of $I_2/I_1$ versus magnetization $\sigma$ for a wavenumber range $0<k<0.01\Omega/c$, showing how for a given $\sigma$, $I_2/I_1$ increases with temperature. For $\sigma\lessapprox1$, the relationship between the two quantities is $I_2/I_1\propto\sigma^{-1/2}$. }
	\label{fig:Isigma}
\end{figure}

The ratio $I_2/I_1$ is presented in Fig. \ref{fig:Isigma}, showing its dependence on both magnetization and temperature. This ratio is equal to $I_2/I_1 = c/v_{A,rel}$ in the linear regime, although the numerical results we show are general. The value of $I_2/I_1$ decreases as the magnetization increases, asymptoting to a value of $I_2/I_1 = 1$ at a temperature dependent magnetization of $\sigma \gtrsim 10$. Physically, this occurs when $v_{A,rel}$ approaches $c$. For lower magnetizations $\sigma \lesssim1$, the relationship becomes $I_2/I_1\propto\sigma^{-1/2}$. This is to be expected as $v_{A,rel} \propto \sigma^{1/2}$.    In this regime, the value of $I_2/I_1$ increases with temperature for all magnetizations in this range due to the decrease in $v_{A,rel}$ at higher temperatures.  

After solving equation \eqref{eq:4} numerically using the above methods, we calculate the fraction of energy contained in the crescent shaped population inversion, $f_{inv}$, as well as the time taken to reach this value. To obtain $f_{inv}$ we take the distribution obtained through solving equation \eqref{eq:4} and calculate its total kinetic energy, $E_{tot} = \int d\textbf{p}(\gamma(\textbf{p})-1)F(\textbf{p})$. We compare this value to the total kinetic energy of another distribution with the same widths as the actual distribution in both the parallel and perpendicular directions. However, this comparison distribution is centred at the coordinates of the maximum value of the actual distribution for all values of $q_\perp$. The resulting distribution is therefore similar to a Bi-Maxwellian distribution with non-constant widths as the contributions from the crescent shaped arms, which are centred at increasingly large values of $q_\parallel>0$ as $|q_\perp|$ increases, have been removed. We denote the total kinetic energy of this modified distribution as $E_{bi}$. $f_{inv}$ is then retrieved by comparing the two energy values through $f_{inv} = (E_{tot}-E_{bi})/E_{tot}$. This quantity is therefore the amount of excess energy contained in the crescent part of the deformed distribution only, and does not include any contributions from changes in energy that are not associated with the population inversion.

\begin{figure*}
		\centering
		\includegraphics[width=2\columnwidth]{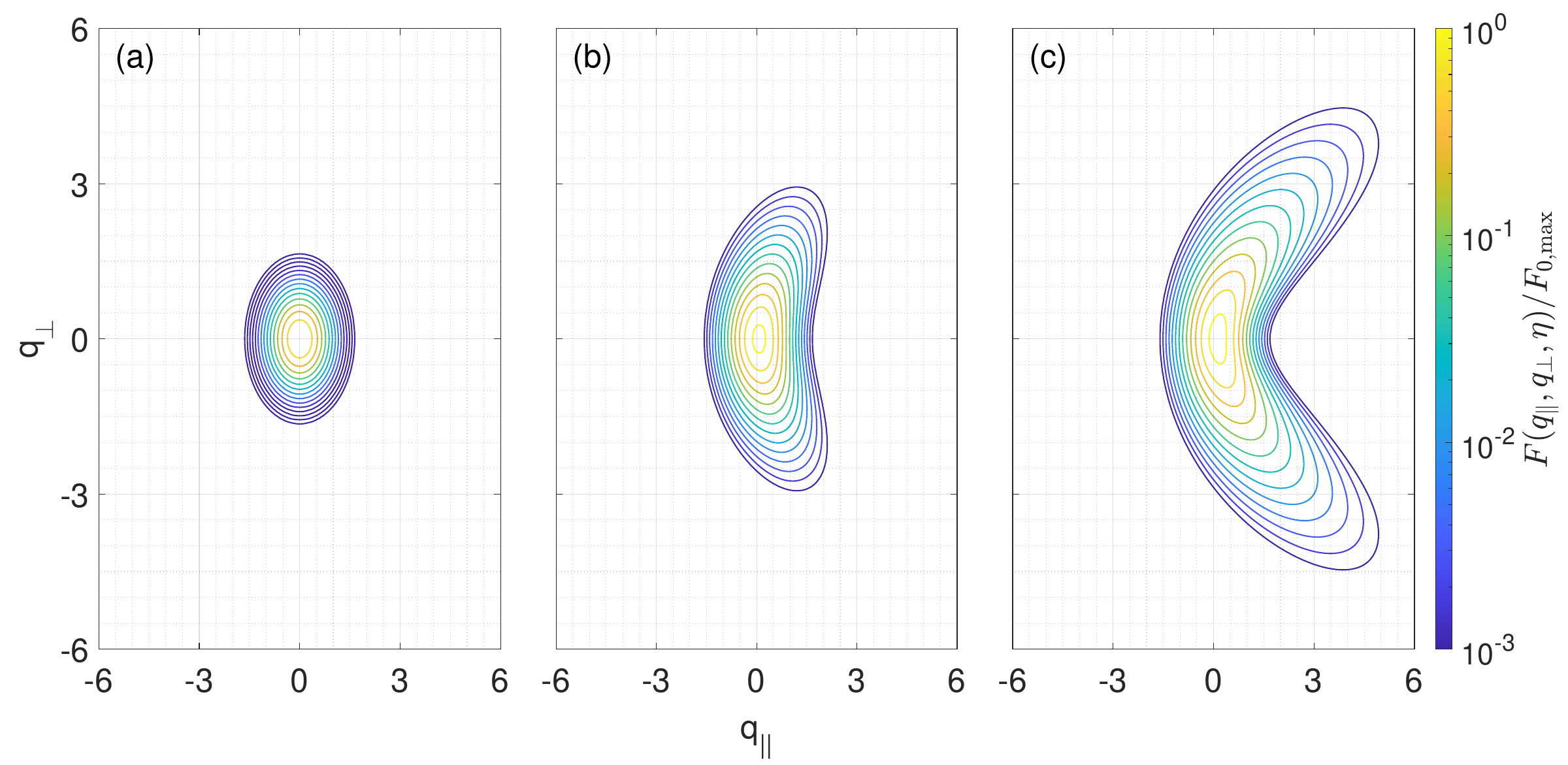}
  {\phantomsubcaption\label{fig:Tm1_t0}%
\phantomsubcaption\label{fig:Tm1_t50}%
\phantomsubcaption\label{fig:Tm1_t500}}
		\caption{Contour plots of the distribution function $F$ for an initial Maxwell-J\"{u}ttner distribution in the high magnetization regime with $\theta = 10^{-1}$ and $I_2/I_1 = 1$. \textbf{Panel (a)} shows the initial distribution function $F_0$; \textbf{panel (b)} shows $F$ at $\eta = 0.3$ and \textbf{panel (c)} shows the distribution at $\eta = 1$. The progression from the initial to the final timestep show the development of the arms of the crescent shaped population inversion.  The values are normalized to the maximum initial value of $F_0(\theta = 0.1)$ and are plotted on a log scale.}
		\label{fig:crescent_plots}
	\end{figure*}

Results were obtained across a large parameter space, examining temperatures ranging from $\theta = 10^{-2}$ to $ \theta = 3$ and magnetizations greater than $\sigma \approx 10^{-4}$. As a demonstration of the evolution of the initial Maxwell-J\"uttner distribution, Fig. \ref{fig:crescent_plots} shows contour plots of the distribution at $\eta =0$, $\eta = 0.3$ and $\eta = 1$, where $\tau_{\max}$ is the time taken to reach the asymptotic value of $f_{inv}$. The results in Fig. \ref{fig:crescent_plots} are for an initial temperature of $\theta = 10^{-1}$ and $I_2/I_1=1$, equivalent to $\sigma\gtrsim10^2$. The formation of the crescent shape is visible through the extension of the initial distribution to larger values of $|q_\perp|$ for $q_\parallel>0$, with the arms of the crescent expanding as the distribution continues to evolve.

\subsection{Detailed analysis of the nonresonant interaction at high magnetizations}
\label{sec:analysis}

\begin{figure}
	 	\centering
	 	\includegraphics[width=\columnwidth]{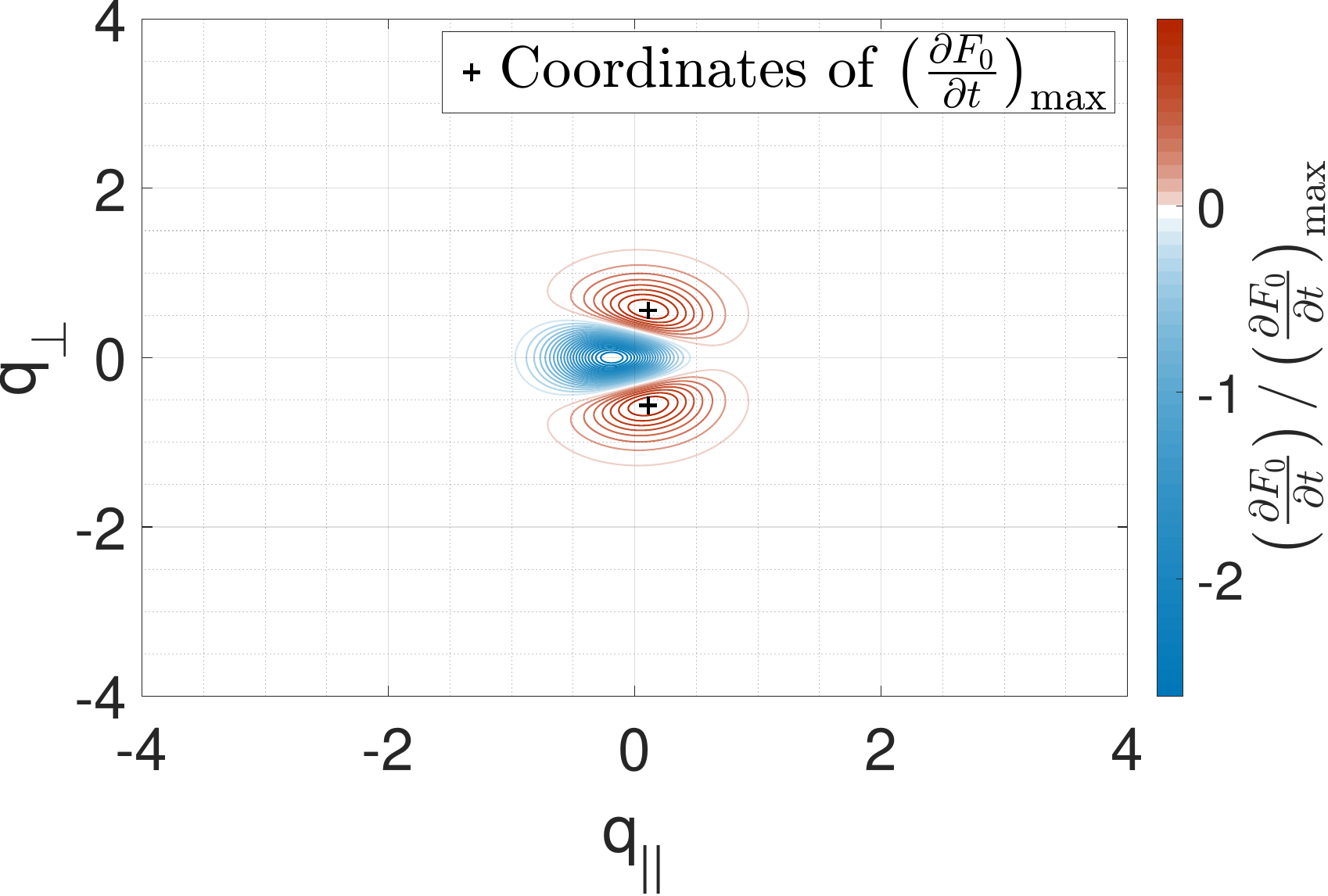}
	 	\caption{Contour plot of $\frac{\partial F_0}{\partial t}$ for a Maxwell-J\"{u}ttner distribution in the high magnetization regime for $\theta = 0.1$ and $I_2/I_1$ = 1. The values are normalized to the maximum value of $\frac{\partial F_0}{\partial t}$. The coordinates of the maximum value of $\frac{\partial F_0}{\partial t}$ are marked by the black crosses. The positive contribution of this term is strongest off-axis in the $q_\parallel>0$ region, and will lead to an evolved distribution as in Fig. \ref{fig:Tm1_t50}. }
	 	\label{fig:dfall}
	 \end{figure}

We first examine the overall evolution of the particle distribution function.  
A plot of $\frac{\partial F_0}{\partial t}$ is shown in Fig. \ref{fig:dfall} for an initial Maxwell-J\"uttner distribution with temperature $\theta = 0.1$ and $I_2/I_1=1$, the same parameters as displayed in Fig. \ref{fig:crescent_plots}. These results are thus applicable for high magnetizations of $\sigma>10^2$. As shown in Fig. \ref{fig:dfall}, the primary outcome of the wave particle interaction is to transfer particles from regions of low perpendicular momentum, $|q_\perp|\lessapprox0.5$, and negative parallel momentum, $q_\parallel <0$ (the blue region in Fig. \ref{fig:dfall}), to higher perpendicular momenta (in both the positive and negative directions) through pitch angle diffusion. The resulting regions of positive $\frac{\partial F}{\partial t}$ are primarily in the $q_\parallel >0$ half of the momentum space. The interaction also results in a bulk momentum in the parallel direction, as can be seen in Fig. \ref{fig:Tm1_t500} where the peak of the distribution is located at $q_\parallel\sim1$. 

\begin{figure*}
	 	\centering
	 	\includegraphics[width=2\columnwidth]{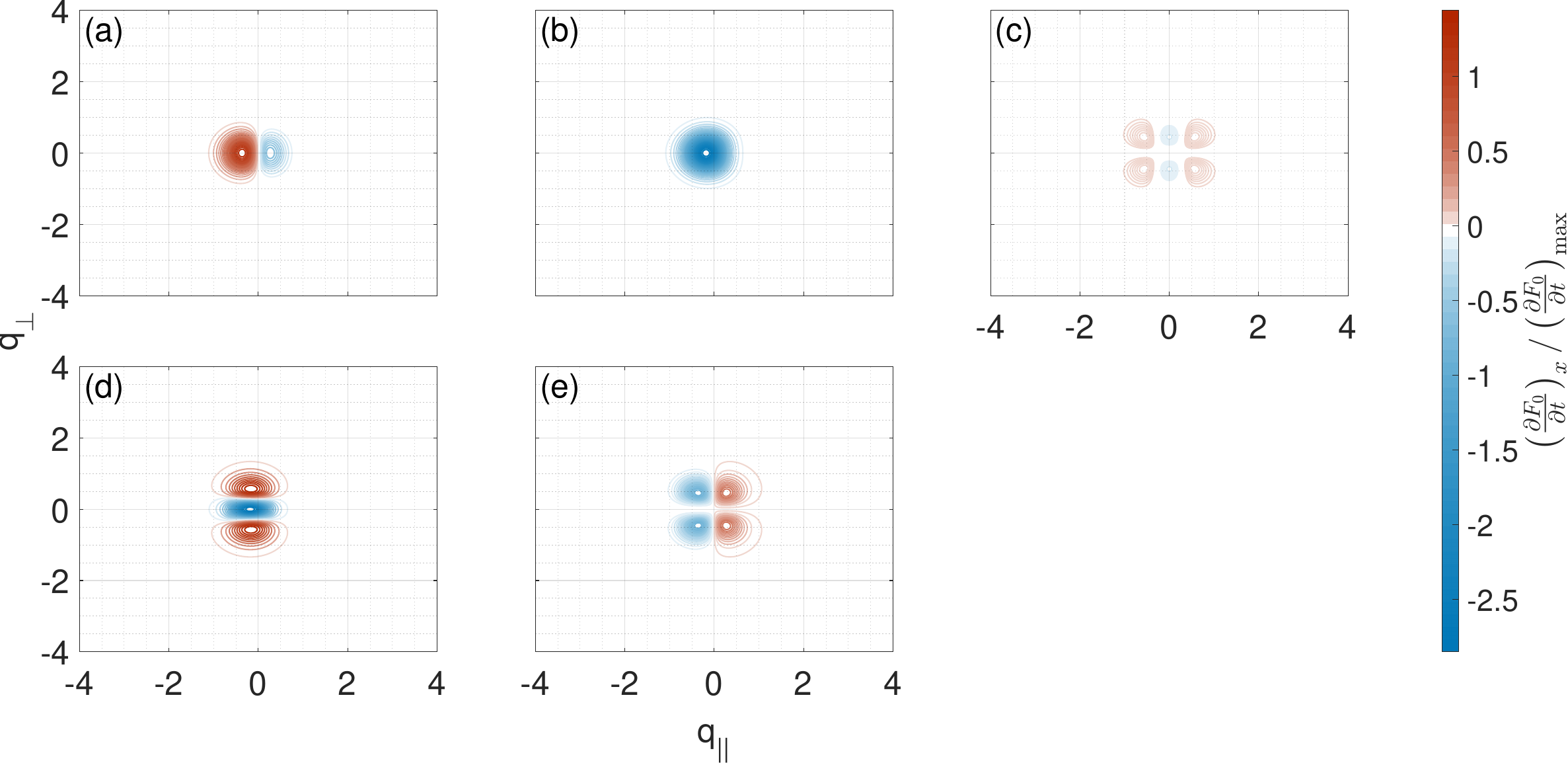}{\phantomsubcaption\label{fig:dfpar1}%
\phantomsubcaption\label{fig:dfperp1}%
\phantomsubcaption\label{fig:dfpar2}%
     \phantomsubcaption\label{fig:dfperp2}%
    \phantomsubcaption\label{fig:dfmix}%
    }

	 	\caption{Contour plots of each individual term $\left(\frac{\partial F_0}{\partial t}\right)_{x}$ from equation \eqref{eq:4}. Each $\left(\frac{\partial F_0}{\partial t}\right)_{x}$ is calculated for a Maxwell-J\"{u}ttner distribution in the high magnetization regime with $\theta = 0.1$ and $I_2/I_1$ = 1. The values are normalized to the maximum value of the overall $\frac{\partial F_0}{\partial t}$. \textbf{Panel (a)} shows the parallel advection term $\left(\frac{\partial F_0}{\partial t}\right)_{\parallel,1}$ The contribution from this term consists of a strong $\frac{\partial F}{\partial t}>0$ for $q_\parallel<0$ and a weaker $\frac{\partial F}{\partial t}<0$ for $q_\parallel>0$. \textbf{Panel (b)} shows the perpendicular advection term $\left(\frac{\partial F_0}{\partial t}\right)_{\perp,1}$. This term has an entirely negative contribution to the change in the distribution for these parameters and thus does not contribute to the formation of the crescent shaped inversion. \textbf{Panel (c)} shows the parallel diffusion term $\left(\frac{\partial F_0}{\partial t}\right)_{\parallel,2}$. The maximum positive contribution for this term is strongest for $|q_\perp|>0$, as in the overall $\frac{\partial F_0}{\partial t}$. However, this term is symmetric in both directions and is also relatively weak compared to the other terms in equation \eqref{eq:4}. \textbf{Panel (d)} shows the perpendicular diffusion term $\left(\frac{\partial F_0}{\partial t}\right)_{\perp,2}$. This term acts to symmetrically diffuse the particles in the perpendicular direction. Note the similarity to  $\frac{\partial F_0}{\partial t}$ in Fig. \ref{fig:dfall}, except for the location of the main regions where $\frac{\partial F_0}{\partial t}>0$. Finally, \textbf{Panel (e)} shows the mixed term $\left(\frac{\partial F_0}{\partial t}\right)_{mix}$. The maximum positive contribution for this term is strongest for $|q_\perp|>0$ and $q_\parallel > 0$, as in the overall $\frac{\partial F_0}{\partial t}$. This term is also considerably stronger than $\left(\frac{\partial F_0}{\partial t}\right)_{\parallel,2}$, which has a similar form in the $q_\parallel>0$ part of the momentum space.}
	 	\label{fig:dfterm}
	 \end{figure*}

To determine how the nonresonant interaction results in the overall evolution shown in Fig. \ref{fig:dfall}, it is necessary to compare the terms in equation \eqref{eq:4} with each other. Examining the contributions to $\frac{\partial F_0}{\partial t}$ from each individual term in Fig. \ref{fig:dfterm} allows us to determine which terms are most crucial to the formation of the population inversion. In our previous work \citep{2023PhRvD.107l1301L}, we focused on the behaviour of the $\left(\frac{\partial F_0}{\partial t}\right)_{\parallel,1}$ term, as it is the largest in magnitude for all values of $I_2/I_1\gtrsim1$. Here, we undertake a more detailed analysis of the relative importance of each term. All five terms are presented in Fig. \ref{fig:dfpar1} to \ref{fig:dfmix}, showing $\left(\frac{\partial F_0}{\partial t}\right)_{\parallel,1}$, $\left(\frac{\partial F_0}{\partial t}\right)_{\perp,1}$, $\left(\frac{\partial F_0}{\partial t}\right)_{\parallel,2}$, $\left(\frac{\partial F_0}{\partial t}\right)_{\perp,2}$ and $\left(\frac{\partial F_0}{\partial t}\right)_{mix}$  respectively for the same set of parameters as presented in Fig. \ref{fig:dfall}. We note that $\frac{\partial F_0}{\partial t}$ is symmetric in the perpendicular direction, but asymmetric in the parallel direction, which is important to the formation of the crescent shape seen in Fig. \ref{fig:crescent_plots}. Indeed, several of the terms in equation \eqref{eq:4} are asymmetric about $q_\parallel = 0$. Both perpendicular terms have a maximum magnitude offset from $q_\parallel = 0$, as can be seen in Fig. \ref{fig:dfperp1} and \ref{fig:dfperp2}. This offset is due to the $I_3$ terms in both coefficients, shown on lines 2 and 4 of equation \eqref{eq:4}.

However, the only two terms where the maximum value of $\frac{\partial F_0}{\partial t}$ is centred at $|q_\perp|>0$ and $q_\parallel >0$ are the parallel diffusion term $\left(\frac{\partial F_0}{\partial t}\right)_{\parallel,2}$, displayed in Fig. \ref{fig:dfpar2}, and the mixed term $\left(\frac{\partial F_0}{\partial t}\right)_{mix}$, shown in Fig. \ref{fig:dfmix}. The parallel diffusion term is considerably lower in magnitude than the mixed term, and furthermore is symmetric in both directions. As none of the other terms provide the necessary positive regions of $\frac{\partial F_0}{\partial t}$ at $|q_\perp|>0$ and $q_\parallel >0$, this leads us to conclude that $\left(\frac{\partial F_0}{\partial t}\right)_{mix}$ is the crucial term that determines the level of population inversion that is formed. This is the case even though it is not the largest term for any set of parameters. The overall form of $\frac{\partial F_0}{\partial t}$ in Fig. \ref{fig:dfall} is dominated by the perpendicular terms, but the contribution of the mixed term provides the element that results in the maximum value of $\frac{\partial F_0}{\partial t}$ being centred at a value of $q_\parallel > 0$ rather than $q_\parallel = 0$. At this high magnetization, the impact of the substantial region of positive $\frac{\partial F_0}{\partial t}$ from the parallel advection term in Fig. \ref{fig:dfpar1} is cancelled out by the equivalent negative $\frac{\partial F_0}{\partial t}$ in the same region from both perpendicular terms and the mixed term.

\subsection{Examining the magnetization and temperature dependence}
\label{sec:alpha}
\begin{figure*}
		\centering
		\includegraphics[width=2\columnwidth]{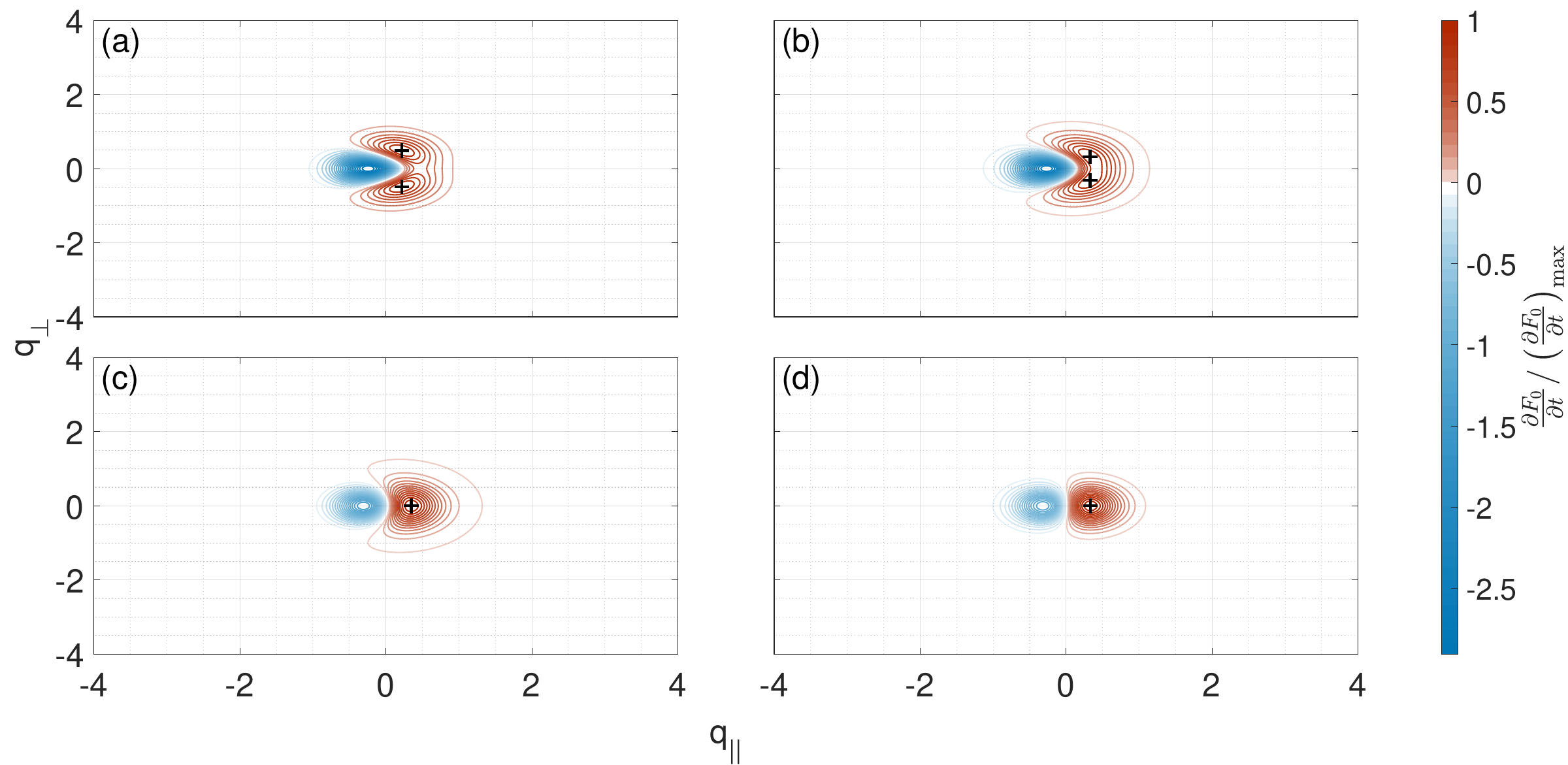}
  {\phantomsubcaption\label{fig:dfalli2}%
\phantomsubcaption\label{fig:dfalli3}%
\phantomsubcaption\label{fig:dfalli10}%
\phantomsubcaption\label{fig:dfalli30}}
		\caption{Contour plots of $\frac{\partial F_0}{\partial t}$ for Maxwell-J\"{u}ttner distributions with $\theta = 0.1$. \textbf{Panel (a)} shows $\frac{\partial F_0}{\partial t}$ for $I_2/I_1 =2$, \textbf{panel (b)} shows $\frac{\partial F_0}{\partial t}$ for $I_2/I_1 =3$, \textbf{panel (c)} shows $\frac{\partial F_0}{\partial t}$ for $I_2/I_1 =10$ and \textbf{panel (d)} shows $\frac{\partial F_0}{\partial t}$ for $I_2/I_1 =30$. In each case the values are normalized to the maximum value of $\frac{\partial F_0}{\partial t}$. The coordinates of the maximum value of $\frac{\partial F_0}{\partial t}$ are marked by the black crosses. The decrease in $\alpha$ with increasing $I_2/I_1$ (equivalent to decreasing magnetization) is visible in the progression from panel (a) to panel (c) and (d), with $\alpha = 0$ for $I_2/I_1 = 10$ and $I_2/I_1 = 30$. }
		\label{fig:iprog}
	\end{figure*}

\begin{figure}
	\centering
	\includegraphics[width=\columnwidth]{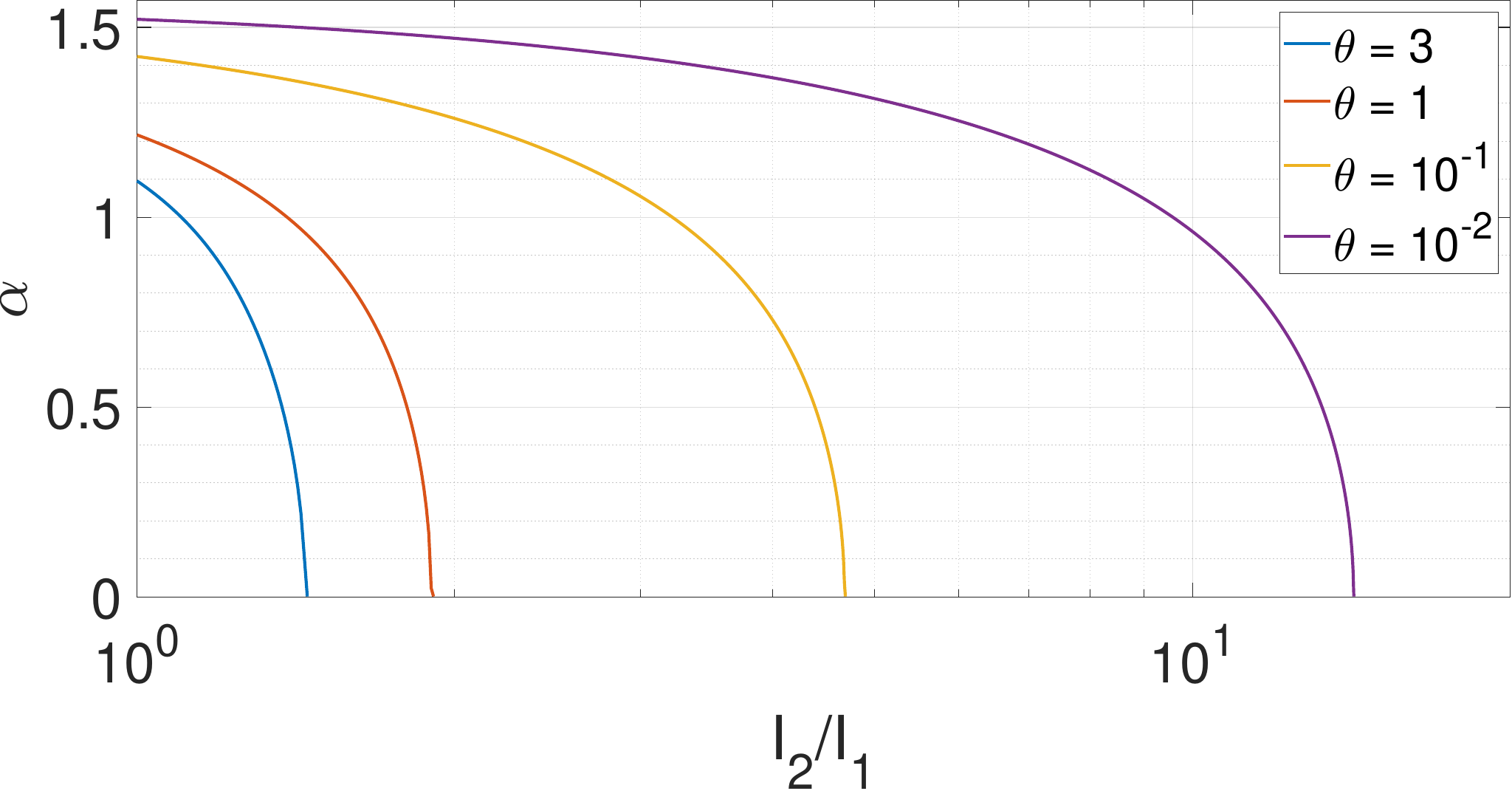}
	\caption{The angle $	\alpha = \tan^{-1}\left(\frac{q_{\perp,\max}}{q_{\parallel,\max}}\right)$ is plotted versus $I_2/I_1$ for temperatures ranging from $\theta = 10^{-2}$ to $\theta = 3$. The solution becomes closer to perpendicular $(\alpha = \pi/2)$ at lower temperatures as well as at higher magnetizations (lower values of $I_2/I_1$). }
	\label{fig:angle}
\end{figure}

To obtain an efficient population inversion with a high value of $f_{inv}$, one requires an evolution of the particle distribution in both the parallel and perpendicular directions. 
The form of $\frac{\partial F_0}{\partial t}$ (i.e. the general regions in momentum space where the change in the distribution function is positive or negative) is strongly dependent on the magnetization. This can be seen in the progression of Fig. \ref{fig:dfall}, \ref{fig:dfalli2}, \ref{fig:dfalli3}, \ref{fig:dfalli10} and \ref{fig:dfalli30} which show $\frac{\partial F_0}{\partial t}$ for $I_2/I_1 = 1$, $I_2/I_1 = 2$, $I_2/I_1 = 3$, $I_2/I_1 = 10$ and $I_2/I_1 = 30$ respectively. The temperature in each case is $\theta = 10^{-1}$. The black crosses in each figure show the coordinates of the maximum value $\left(\frac{\partial F_0}{\partial t}\right)_{\max}$. As the magnetization decreases and $I_2/I_1$ increases, the main positive regions of $\frac{\partial F_0}{\partial t}$ move from being predominantly in the perpendicular direction to the positive parallel direction. This evolution can be quantified by the coordinates of $\left(\frac{\partial F_0}{\partial t}\right)_{\max}$, denoted as $q_{\parallel,\max}$ and $q_{\perp,\max}$. The angle this point makes with the $q_\parallel$ axis is defined as $\alpha = \tan^{-1}\left(\frac{q_{\perp,\max}}{q_{\parallel,\max}}\right)$. This angle is presented in Fig. \ref{fig:angle} as a function of the magnetization for a range of temperatures.

The angle $\alpha$ provides a description of the general form of $\frac{\partial F_0}{\partial t}$. The maximal value of the population inversion occurs for intermediate values of $0 < \alpha < \pi/2$ where a clear crescent shape can form. However, some inversion will occur even for extreme values of $\alpha = 0, \pi/2$, due to the contribution from the mixed term. As discussed in section \ref{sec:analysis}, this provides the crucial positive regions of  $\frac{\partial F}{\partial t}$  for $|q_\perp|>0$ and $q_\parallel >0$. The importance of $\left(\frac{\partial F_0}{\partial t}\right)_{mix}$ for different parameters is further discussed in section \ref{sec:mixed} below.

Fig. \ref{fig:angle} shows the evolution of $\alpha$ for different particle temperatures. While the general form of $\frac{\partial F_0}{\partial t}$ is qualitatively the same for different temperatures, the precise location of the positive and negative regions is temperature dependent. 
As can be seen, $\alpha$ increases towards $\frac{\pi}{2}$ as the temperature decreases, implying  that the distribution function changes predominantly in the perpendicular direction, as described in previous works \citep[e.g.][]{Wu2007,Yoon2009}. While for all parameters there are effects in both the parallel and perpendicular directions, the acceleration of particles in the perpendicular direction through pitch angle diffusion becomes stronger relative to the parallel processes at lower temperatures. This will affect both the angle $\alpha$, as well as the widths of the particle distribution in the parallel and perpendicular directions, $\theta_\parallel$ and $\theta_\perp$ \footnote{ Note that at $t=0$, $\theta_\parallel = \theta_\perp = \theta$, but they can differ with time.}. The results for the temperature dependence of $\alpha$ obtained match expectations from previous analyses in the non-relativistic regime which show that the temperature anisotropy $A=\theta_\perp/\theta_\parallel$ caused by the nonresonant interaction decreases as the temperature increases \citep{2006PhRvL..96l5001W}.

\begin{figure}
	 	\centering
	 	\includegraphics[width=\columnwidth]{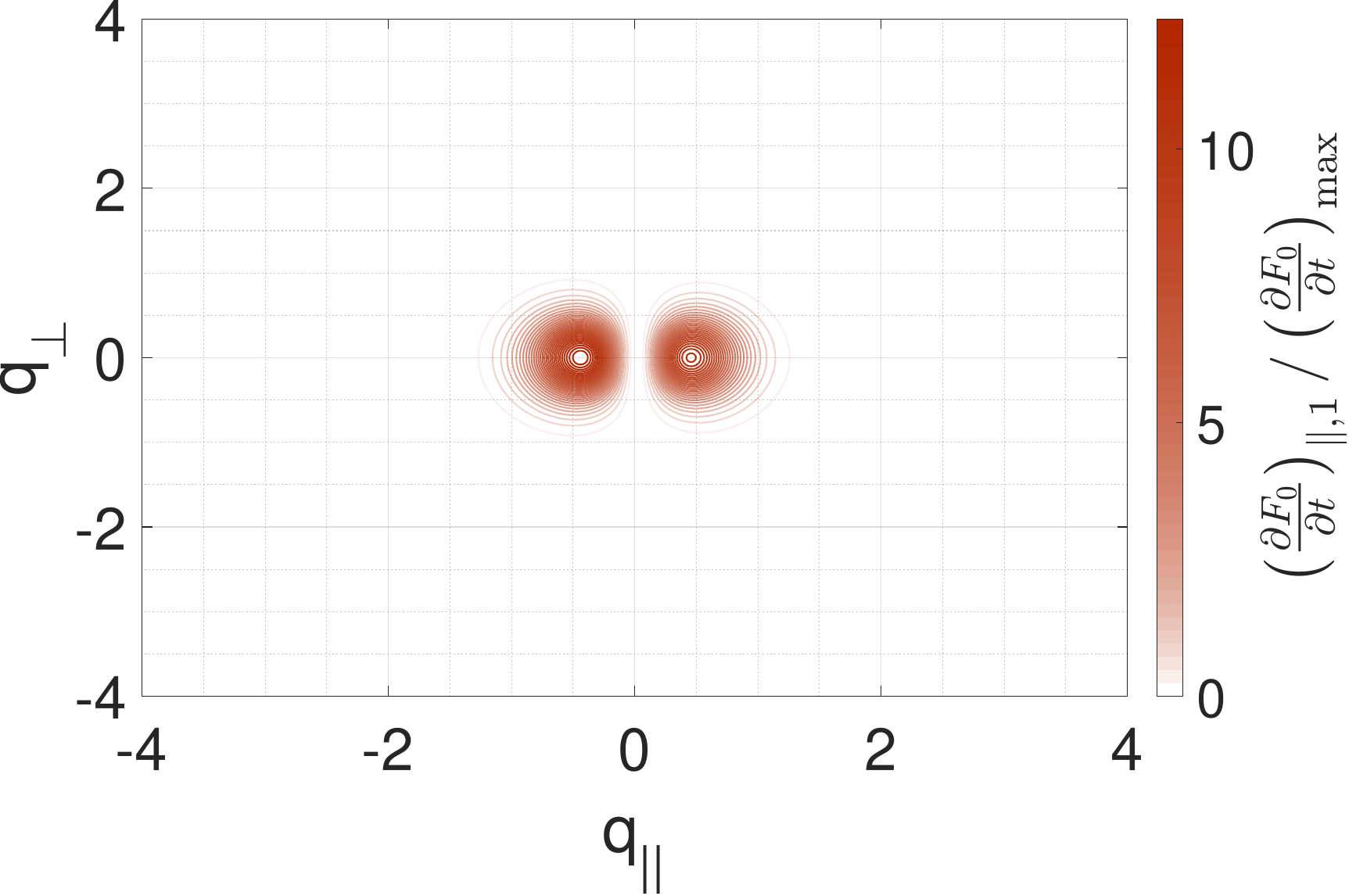}
	 	\caption{Contour plot of $\left(\frac{\partial F_0}{\partial t}\right)_{\parallel,1}$ for a Maxwell-J\"{u}ttner distribution in the low/intermediate magnetization regime for $\theta = 0.1$ and $I_2/I_1 = 30$. The values are normalized to the maximum value of $\frac{\partial F_0}{\partial t}$. In contrast to the case in Fig. \ref{fig:dfpar1} where $I_2/I_1=1$, for these parameters the contribution from this term consists of a strong $\left(\frac{\partial F_0}{\partial t}\right)>0$ for both positive and negative $q_\parallel$.}
	 	\label{fig:dfpar1i100}
	 \end{figure}

For all temperatures $\theta$ examined, $\alpha$ always decreases toward 0 as the magnetization decreases. This decrease is largely due to the behaviour of the parallel advection term $\left(\frac{\partial F_0}{\partial t}\right)_{\parallel,1}$. First, as $I_2/I_1$ increases, this term becomes positive for $q_\parallel>0$, which is in contrast to the case for high magnetizations ($I_2/I_1\sim1$) where $\left(\frac{\partial F_0}{\partial t}\right)_{\parallel,1}$ is negative for $q_\parallel>0$. This can be seen by examining the coefficient for this term on line 1 of equation \eqref{eq:4}. As $I_2/I_1$ increases, the term $2\frac{I_2 q_\parallel}{\gamma^2}$ becomes greater in magnitude than the other terms contributing to $\left(\frac{\partial F_0}{\partial t}\right)_{\parallel,1}$. As there is an additional factor of $-q_\parallel$ from the partial derivative $\frac{\partial F}{\partial q_\parallel}$ in the case of a Maxwell-J\"uttner distribution, the dominance of this term results in $\left(\frac{\partial F_0}{\partial t}\right)_{\parallel,1}$ being positive across all of momentum space. This is shown in Fig. \ref{fig:dfpar1i100}, which displays $\left(\frac{\partial F_0}{\partial t}\right)_{\parallel,1}$ for $I_2/I_1=30$. This figure also shows that
 $\left(\frac{\partial F_0}{\partial t}\right)_{\parallel,1}$ is centred at $q_\perp = 0$, contributing to a decrease in $\alpha$. This is in contrast to the contribution from the parallel and mixed diffusion terms, which are not centred at $q_\perp = 0$ (see Fig. \ref{fig:dfpar2} and \ref{fig:dfmix} respectively).

\begin{figure}
		\centering
		\includegraphics[width=\columnwidth]{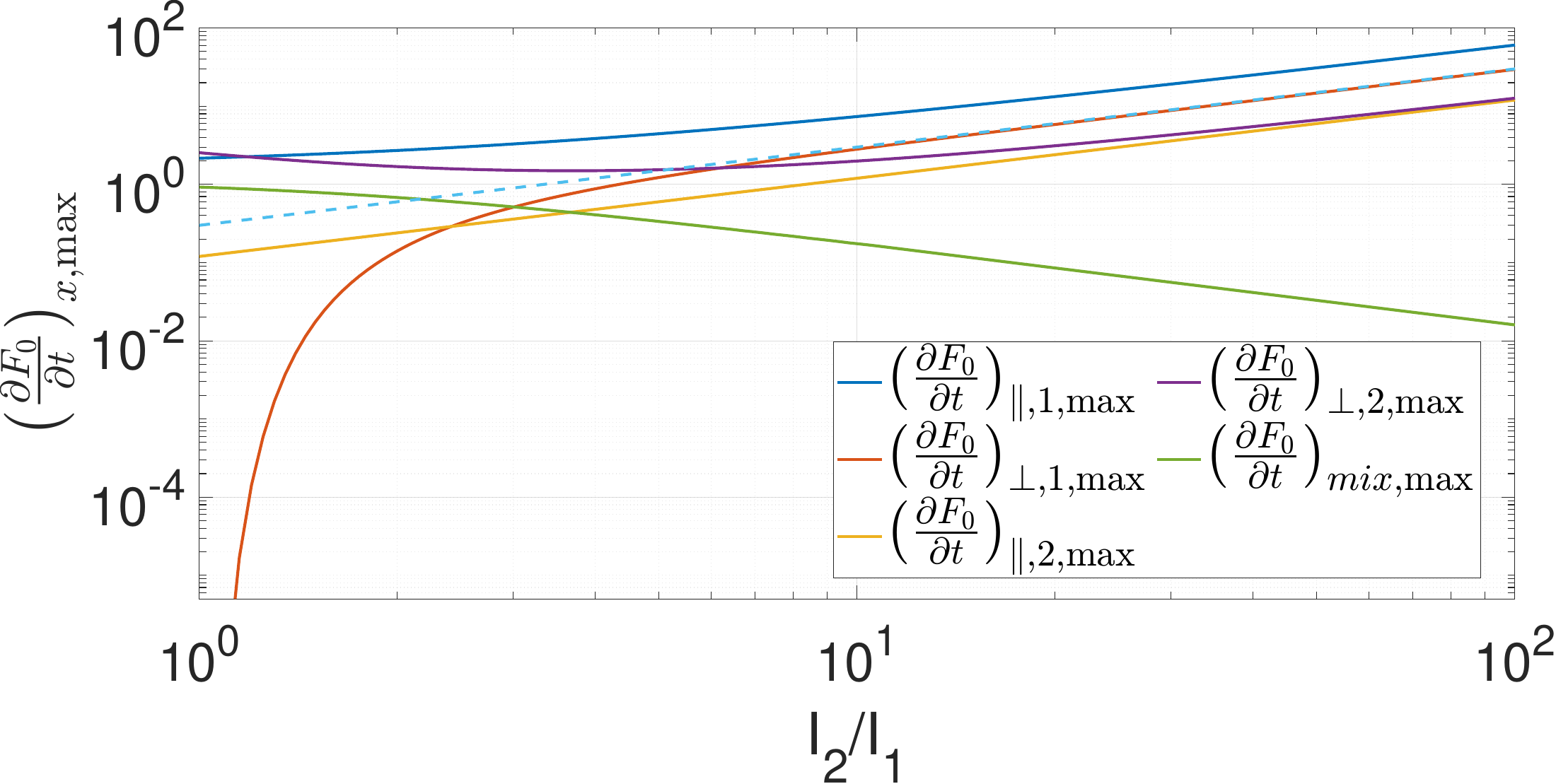}
		\caption{Plot of the the maximum value of every term in equation \eqref{eq:4}, normalized to the maximum value of the total $(\frac{\partial F_0}{\partial t})_{\max}$. The dashed cyan line shows a fit of $\left(\frac{\partial F_0}{\partial t}\right)_{\perp,1,max}\propto I_2/I_1$. This trend is followed by all terms except the mixed one for $I_2/I_1>>1$. These values are calculated for a temperature of $\theta = 10^{-1}$. As the temperature increases, the first order terms $\left(\frac{\partial F_0}{\partial t}\right)_{\parallel,1}$ and $\left(\frac{\partial F_0}{\partial t}\right)_{\perp,1}$ will become stronger relative to the others, though the overall trends will be unchanged.}
		\label{fig:termcomp}
	\end{figure}

Furthermore, the parallel advection term increases in magnitude as $I_2/I_1$ increases, becoming larger than all others at $I_2/I_1\sim1.2$ for $\theta = 10^{-1}$. This can be seen in Fig. \ref{fig:termcomp} where the maximum value of each term $\left(\frac{\partial F_0}{\partial t}\right)_{x,max}$ from equation \eqref{eq:4} is plotted, with $\left(\frac{\partial F_0}{\partial t}\right)_{\parallel,1,\max}$ shown by the solid blue line. These values in Fig. \ref{fig:termcomp} are normalized to the overall $\left(\frac{\partial F_0}{\partial t}\right)_{max}$. The value of $I_2/I_1$ at which the parallel advection term becomes largest is temperature dependent, with the value increasing as the temperature decreases. The other major terms at $I_2/I_1\sim1$ are $\left(\frac{\partial F_0}{\partial t}\right)_{\perp,2}$ and $\left(\frac{\partial F_0}{\partial t}\right)_{mix}$, shown by the solid purple and green lines in Fig. \ref{fig:termcomp} respectively. As both these terms are second order, they are relatively stronger at lower temperatures due to the factor of $\theta^{-2}$ from the derivatives $\frac{\partial^2 F}{\partial q_\perp^2}$ and $\frac{\partial^2 F}{\partial q_\parallel \partial q_\perp}$.
This implies that the ratio $I_2/I_1$ above which the parallel advection term becomes dominant increases at lower temperatures.
The increase in strength of the parallel advection term, as well as the relative weakness of the mixed term as the magnetization decreases results in $\left(\frac{\partial F_0}{\partial t}\right)_{max}$ shifting closer to $|q_\perp| = 0$.  As is shown in Fig. \ref{fig:dfalli10} and \ref{fig:dfalli30} , this combination of factors results in an overall initial evolution that is closer to parallel advection in the low magnetization parameter space, though pitch angle diffusion is still occurring on a smaller scale.
 
The reduction in the relative strength of the pitch angle diffusion relative to the parallel processes when $I_2/I_1>>1$ results in evolved distributions that do not display the obvious crescent shapes observed at high magnetizations (such as in Fig. \ref{fig:Tm1_t500}), as discussed in our previous work \citep {2023PhRvD.107l1301L}. However, particles do still continue to be moved to higher perpendicular momenta for $q_\parallel > 0$ even when $\alpha = 0$. As a result, the population inversion still can form, though with much reduced values of $f_{inv}$, as is shown in our numerical results presented in Section \ref{sec:energy}.

To summarize the conclusions of the previous paragraphs, the acceleration of particles to larger perpendicular momenta values becomes relatively stronger both as the magnetization increases and the temperature decreases, resulting in an increase in $\alpha$ up to a maximum value of $\alpha = \pi/2$. This value however does not correspond to the most efficient population inversion as the resulting evolution is almost entirely perpendicular, while for the formation of the population inversion some contribution from the mixed and parallel terms is naturally required to produce a crescent shape. The maximum value of $f_{inv}$ is therefore obtained in the range $0<\alpha<\pi/2$. We note that these results are the equivalent of the mechanism becoming less efficient as the plasma $\beta$ increases, a result also seen in the non-relativistic regime \citep{2009PhPl...16b0703W}. We also emphasize that the maximum value of $f_{inv}$ is not solely dependent on $\alpha$, but also strongly depends on the mixed term as discussed in detail in Sections \ref{sec:analysis} and \ref{sec:mixed}.

\subsection{The mixed term and formation of a population inversion} 
 \label{sec:mixed}
\begin{figure}
		\centering
		\includegraphics[width=\columnwidth]{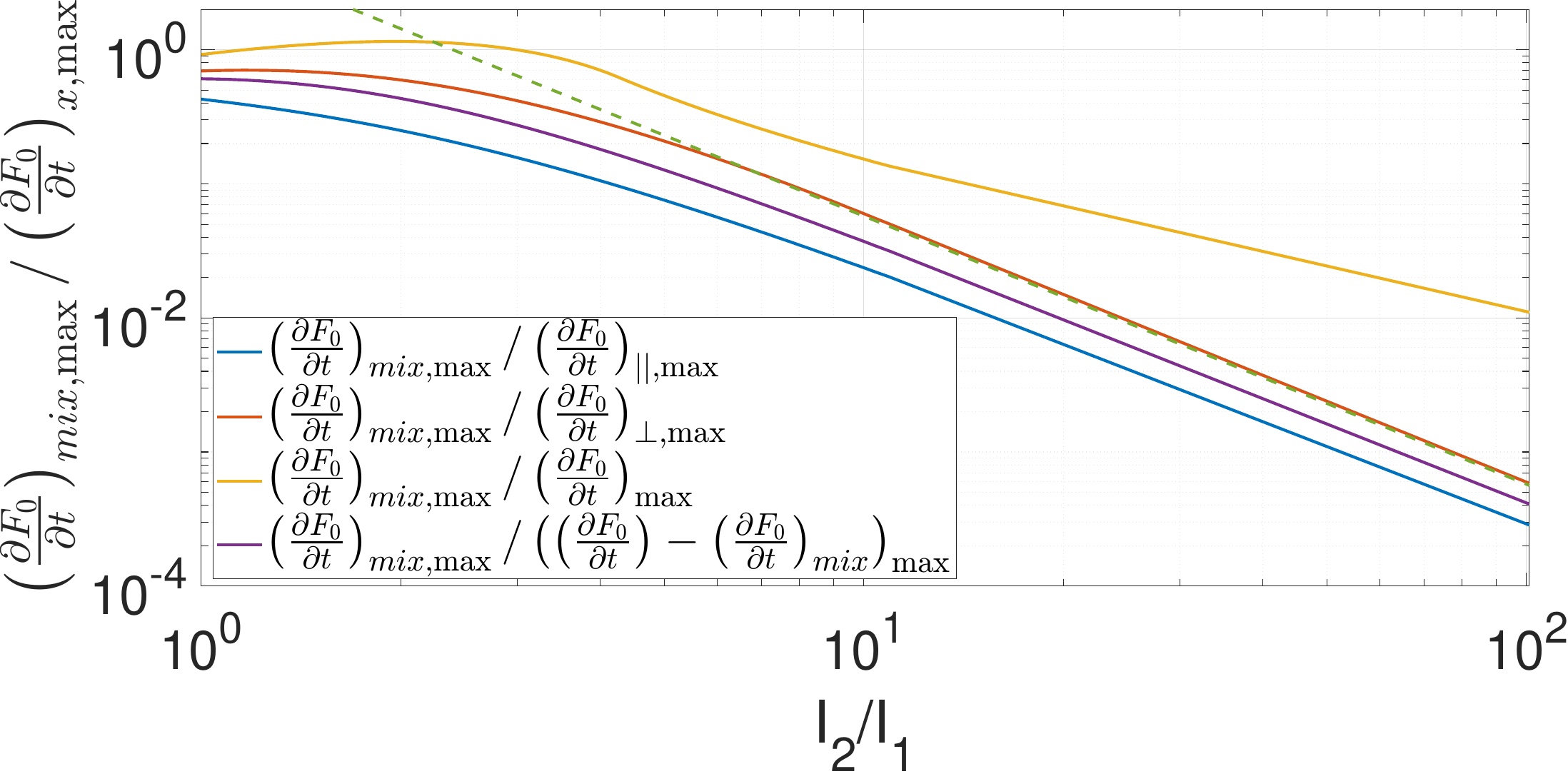}
		\caption{Plot of the ratio of $\left(\frac{\partial F_0}{\partial t}\right)_{mix,\max}$ for the mixed term to the maximum value of the parallel terms  $(\frac{\partial F_0}{\partial t})_{\parallel,\max}$ (blue), perpendicular terms $(\frac{\partial F_0}{\partial t})_{\perp,\max}$ (orange), overall solution $(\frac{\partial F_0}{\partial t})_{\max}$ (yellow) and  $\left(\left(\frac{\partial F_0}{\partial t}\right)- (\frac{\partial F_0}{\partial t})\right)_{\max}$ (purple). The dashed green line indicates a fit of $\left(\frac{\partial F_0}{\partial t}\right)_{mix,\max}/ \left(\frac{\partial F_0}{\partial t}\right)_{\perp,\max}\propto \left(I_2/I_1\right)^{-2}$. These values are calculated for a temperature of $\theta = 10^{-1}$.}
		\label{fig:ratioT1}
	\end{figure}

Based on our previous analysis in section \ref{sec:analysis}, the mixed term is a crucial element to the formation of the population inversion. We emphasize that the mixed term in general is not the largest in magnitude and as such does not dominate the overall form of the evolved distribution, especially at lower magnetizations. However, it provides the necessary positive $\frac{\partial F_0}{\partial t}$ for $|q_\perp|>0$ and $q_\parallel > 0$ that leads to the crescent shape of the distribution. Furthermore, we show in Section \ref{sec:num} below that the trends followed by the mixed term closely match our numerical results. In order to examine the parametric dependence of this term, it is useful to compare its maximum value $\left(\frac{\partial F_0}{\partial t}\right)_{mix,\max}$ to that of the other terms. Fig. \ref{fig:ratioT1} presents the ratio of $\left(\frac{\partial F_0}{\partial t}\right)_{mix,\max}$ to the parallel and perpendicular terms $\left(\frac{\partial F_0}{\partial t}\right)_{\parallel,\max}$ and $\left(\frac{\partial F_0}{\partial t}\right)_{\perp,\max}$ as well to the overall $\left(\frac{\partial F_0}{\partial t}\right)_{\max}$ and the sum of all the other terms except the mixed term $\left(\frac{\partial F_0}{\partial t} - \left(\frac{\partial F_0}{\partial t}\right)_{mix}\right)_{\max}$ . Here $\left(\frac{\partial F_0}{\partial t}\right)_{\parallel,\max} = \left(\left(\frac{\partial F_0}{\partial t}\right)_{\parallel,1}+\left(\frac{\partial F_0}{\partial t}\right)_{\parallel,2}\right)_{max}$ is the maximum value of the combined parallel terms and $\left(\frac{\partial F_0}{\partial t}\right)_{\perp,\max} = \left(\left(\frac{\partial F_0}{\partial t}\right)_{\perp,1}+\left(\frac{\partial F_0}{\partial t}\right)_{\perp,2}\right)_{max}$ is the maximum value of combined perpendicular terms. We combine these terms for clarity of presentation as all the individual terms follow the same trends for $I_2/I_1>>1$, as shown in Fig. \ref{fig:termcomp}. 
  
In all cases except that of the comparison to the overall $\left(\frac{\partial F_0}{\partial t}\right)_{\max}$, the ratio is proportional to $(I_2/I_1)^{-2}$ at large values of $I_2/I_1$ ($I_2/I_1\gtrapprox 6$ for $\theta = 0.1$ as in Fig. \ref{fig:ratioT1}). This is the equivalent of $\left(\frac{\partial F_0}{\partial t}\right)_{mix,\max}/\left(\frac{\partial F_0}{\partial t}\right)_{x,\max} \propto \sigma$  at these low and intermediate magnetizations, as $I_2/I_1\propto \sigma^{-1/2}$ for $\sigma\ll1$. As $I_2/I_1$ approaches 1, equivalent to higher magnetizations of $\sigma \gtrapprox 1$, the ratio flattens out, becoming more weakly dependent on $I_2/I_1$ (see Fig. \ref{fig:Isigma} for the temperature dependent $I_2/I_1$-$\sigma$ relationship). These results are derived semi-analytically in Appendix A below.

\begin{figure}
	\centering
	\includegraphics[width=\columnwidth]{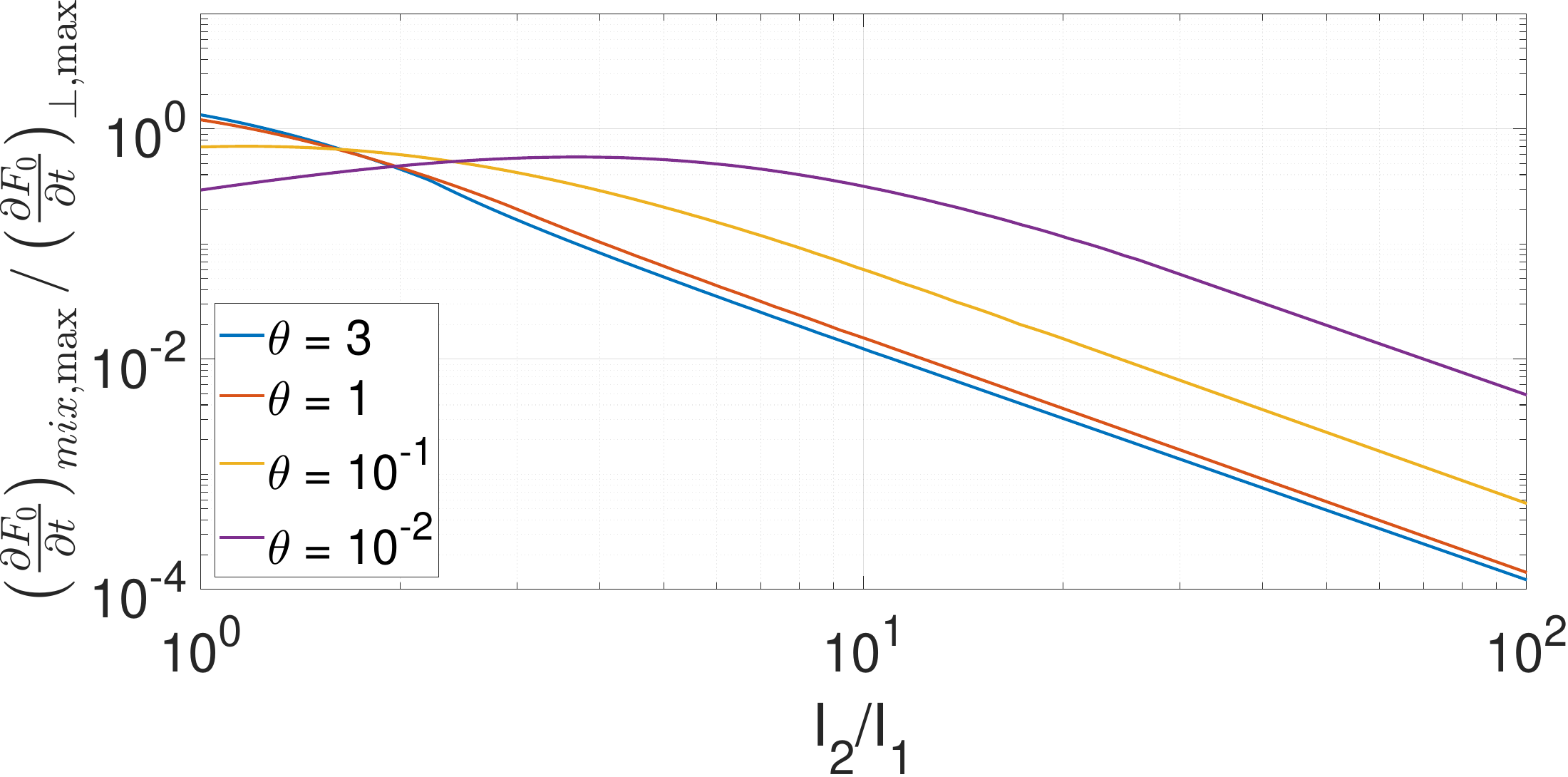}
	\caption{The ratio of the maximum value of the mixed term, $\left(\frac{\partial F_0}{\partial t}\right)_{mix,\max}$, to the maximum value of the perpendicular terms, $\left(\frac{\partial F_0}{\partial t}\right)_{\perp,\max}$, plotted versus $I_2/I_1$. Each line represents a different temperature, with the peak of the ratio not at $I_2/I_1 = 1$ for $\theta = 10^{-2}$, as discussed in Section \ref{sec:num}. }
	\label{fig:f4f5}
\end{figure}

These trends are true for all temperatures, though the magnetization at which the linear relationship begins is temperature dependent. This can be seen in Fig. \ref{fig:f4f5}, which presents the ratio $\left(\frac{\partial F_0}{\partial t}\right)_{mix,\max}/\left(\frac{\partial F_0}{\partial t}\right)_{\perp,\max}$ for different temperatures. The choice of this specific ratio in Fig. \ref{fig:f4f5} and in Section \ref{sec:num} below is because the other ratios $\left(\frac{\partial F_0}{\partial t}\right)_{mix,\max}/\left(\frac{\partial F_0}{\partial t}\right)_{\parallel,\max}$ and $\left(\frac{\partial F_0}{\partial t}\right)_{mix,\max}/\left(\frac{\partial F_0}{\partial t} - \left(\frac{\partial F_0}{\partial t}\right)_{mix}\right)_{\max}$ show the same trends, and thus carry no new information. An increase in the value of $I_2/I_1$ at which the $\left(\frac{\partial F_0}{\partial t}\right)_{mix,\max}/\left(\frac{\partial F_0}{\partial t}\right)_{\perp,\max}\propto(I_2/I_1)^{-2}$ relationship begins can be seen as the temperature decreases.

\begin{figure}
	\centering
	\includegraphics[width=\columnwidth]{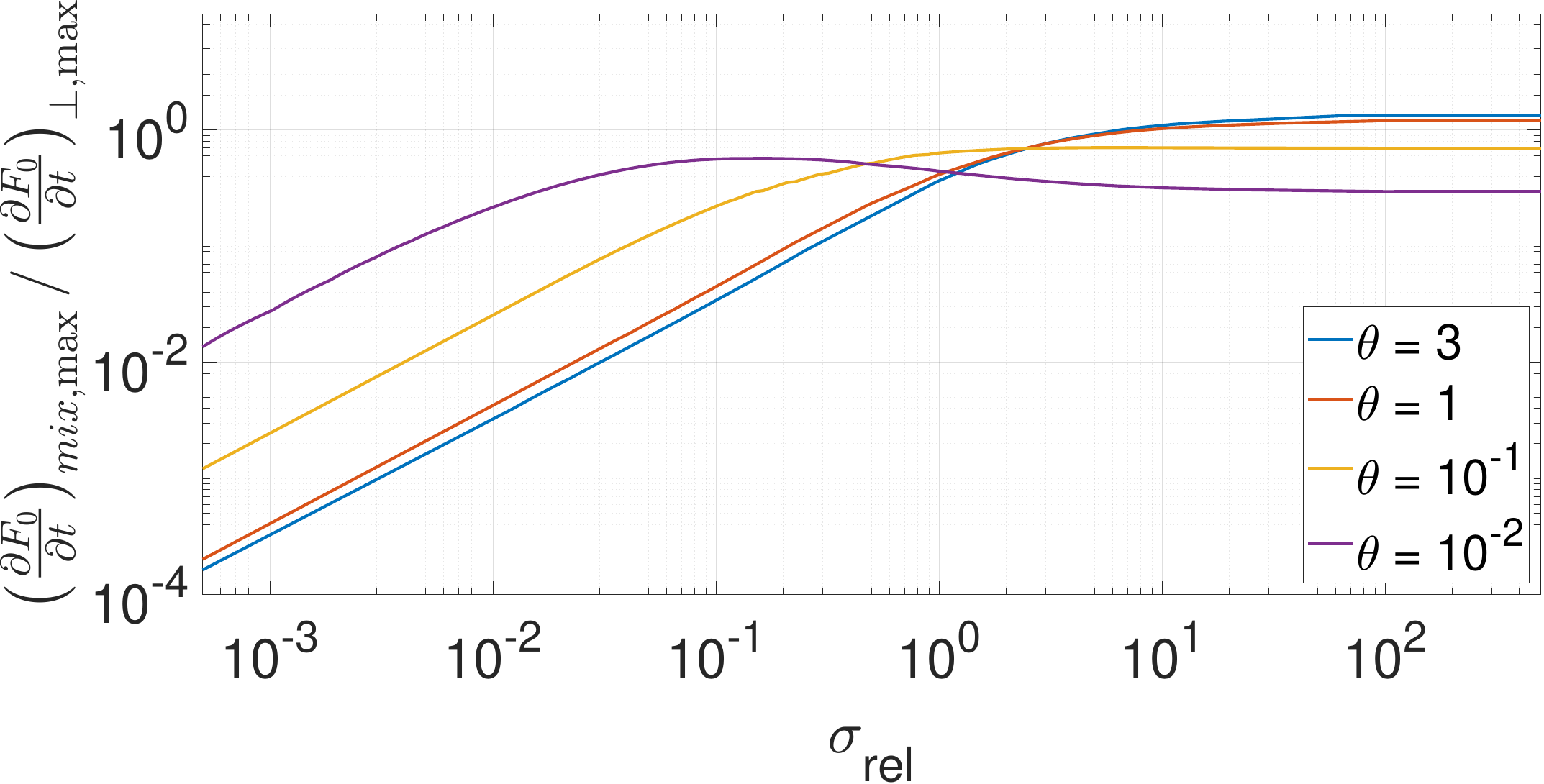}
	\caption{The ratio of the maximum value of the mixed term, $\left(\frac{\partial F_0}{\partial t}\right)_{mix,\max}$, to the maximum value of the perpendicular terms, $\left(\frac{\partial F_0}{\partial t}\right)_{\perp,\max}$, plotted versus the relativistic magnetization $\sigma_{rel}$. Note the broadness of the region where  $\left(\frac{\partial F_0}{\partial t}\right)_{mix,\max}/\left(\frac{\partial F_0}{\partial t}\right)_{\perp,\max}$ is not proportional to $\sigma_{rel}$ in comparison to the equivalent region in Fig. \ref{fig:f4f5}, caused by the asymptotic behaviour of $\sigma$ as a function of $I_2/I_1$ for $\sigma>1$.}
	\label{fig:f4f5sig}
\end{figure}

To clarify this behaviour in terms of a more intuitive physical quantity, the same ratio is plotted in Fig. \ref{fig:f4f5sig} versus the relativistic magnetization $\sigma_{rel} = \sigma/\gamma_{av}$, where $\gamma_{av}$ is the average Lorentz factor of the initial distribution. This shows the $\left(\frac{\partial F_0}{\partial t}\right)_{mix,\max}/\left(\frac{\partial F_0}{\partial t}\right)_{\perp,\max} \propto \sigma_{rel}$ trend at low magnetizations as described above. The region where the ratio flattens out, namely $I_2/I_1\lesssim 2-10$ (depending on the temperature), corresponds to a much larger region in terms of the magnetization due to the asymptotic behaviour of the $I_2/I_1-\sigma$ relationship as the relativistic Alfv\'en velocity approaches the speed of light. This results in the relatively broader non-power law regions in Fig. \ref{fig:f4f5sig} compared to Fig. \ref{fig:f4f5}. These trends are reproduced in the numerical results for $f_{inv}$ presented below in Section \ref{sec:num}, supporting our conclusion about the importance of the mixed term.

\section{Results}
\label{sec:num}
\subsection{The energy in the population inversion}
\label{sec:energy}
\begin{figure*}
		\centering
		\includegraphics[width=2\columnwidth]{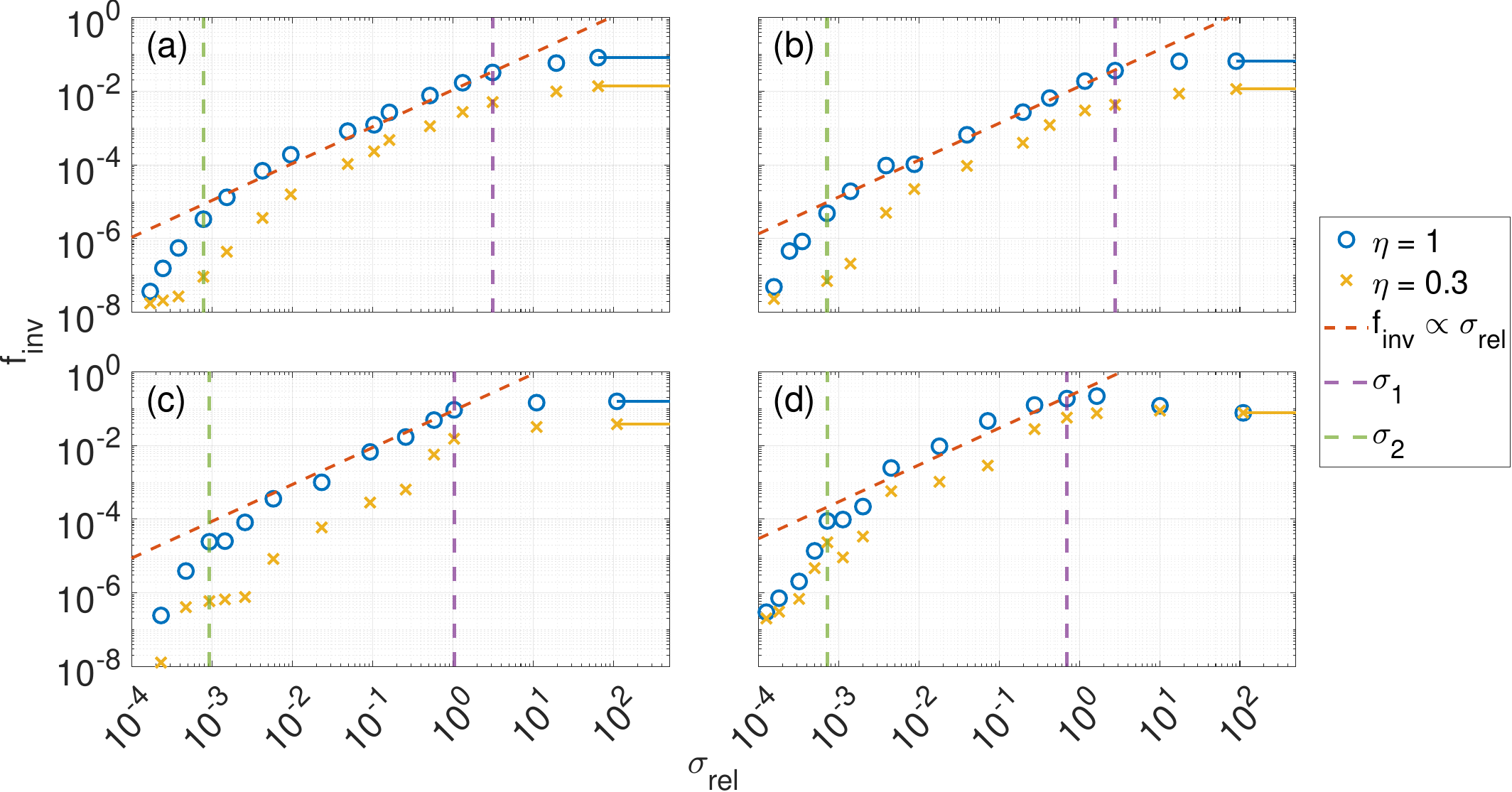}
  {\phantomsubcaption\label{fig:T3}%
\phantomsubcaption\label{fig:T1}%
\phantomsubcaption\label{fig:Tm1}%
\phantomsubcaption\label{fig:Tm2}}
		\caption{The energy in the inversion versus magnetization, with different temperatures in each panel. \textbf{Panel (a)} presents results for $\theta = 3$, \textbf{panel (b)} for $\theta = 1$, \textbf{panel (c)} for $\theta = 10^{-1}$ and \textbf{panel (d)} shows results for $\theta = 10^{-2}$. In each panel the blue circles show the energy fraction obtained for $\eta = 1$, while the yellow 'x's show the energy fraction at $\eta = 0.3$. The purple dashed line ($\sigma_1$) divides the high and intermediate magnetization regime, and the green dashed line ($\sigma_2$) is the boundary between the low and intermediate magnetization regimes, showing where the cut-off occurs. The dashed orange line shows a fit of $f_{inv} \propto \sigma_{rel}$ in the intermediate magnetization regime in each case.}
		\label{fig:energies}
	\end{figure*}

Fig. \ref{fig:T3} to \ref{fig:Tm2} show the fraction of energy in the population inversion, $f_{inv}$ versus $\sigma_{rel}$ for temperatures of $\theta =\lbrace 3, 1, 10^{-1},10^{-2}\rbrace$ respectively. All the results below are presented in terms of $\sigma_{rel}$ rather than $I_2/I_1$ as the magnetization is a more useful physical quantity. The results are presented for values of $\eta = 1$ and $\eta = 0.3$. This range of values represents the upper limits on the maximum achievable values of $f_{inv}$ while satisfying the constraints of quasilinear theory. The theory is valid for values of $\eta$ not much smaller than unity \citep[e.g.][]{Yoon2009}. Magnetizations of $\sigma_{rel}>10^{-4}$ were examined. The results show that the behaviour of $f_{inv}$ with respect to the magnetization of the plasma can be divided up into three regimes. These consist of a high magnetization regime where $f_{inv}$ is only weakly dependent on $\sigma_{rel}$, an intermediate magnetization regime where $f_{inv} \propto \sigma_{rel}$,  and a low magnetization regime where $f_{inv}$ decreases rapidly as the magnetization decreases. These regimes and their properties are discussed in detail below, and a summary of the different regimes is given in Table \ref{tab:regimes}.

The transition between the high and intermediate magnetization regimes occurs at $\sigma_{rel} = \sigma_1$. We define this value as the maximum magnetization for which $f_{inv}\propto \sigma_{rel}$.  Above this value, the fraction of energy in the population inversion depends weakly on the magnetization, as can be seen by the flattening out of $f_{inv}$ for $\sigma_{rel} > \sigma_1$ in Fig. \ref{fig:T3} to \ref{fig:Tm2}. This transition occurs for all $\eta$ values presented. We note that this transition is not a sharp cut-off at precisely $\sigma_1$, but a more gradual change, as can be seen in both the numerical results (Fig. \ref{fig:T3} to \ref{fig:Tm2}) and the ratio of $\left(\frac{\partial F_0}{\partial t}\right)_{mix,\max}/\left(\frac{\partial F_0}{\partial t}\right)_{\perp,\max}$ (Fig. \ref{fig:f4f5} and \ref{fig:f4f5sig}). As discussed in Section \ref{sec:analysis}, the relative strength of the mixed term is important to the formation of the population inversion. Indeed, the trend of $f_{inv}$ can be explained by the similar behaviour of $\left(\frac{\partial F_0}{\partial t}\right)_{mix,\max}/\left(\frac{\partial F_0}{\partial t}\right)_{\perp,\max}$ in the same magnetization regime, as shown in Fig. \ref{fig:f4f5sig}.

While for higher temperatures of $\theta > 10^{-1}$ the fraction of energy in the inversion simply transitions to an asymptotic value for $\sigma_{rel} > \sigma_1$, this is not the case for $\theta = 10^{-2}$ (see Fig. \ref{fig:Tm2}). At this lower temperature, both $f_{inv}$ and $\left(\frac{\partial F_0}{\partial t}\right)_{mix,\max}/\left(\frac{\partial F_0}{\partial t}\right)_{\perp,\max}$ initially increase with $\sigma_{rel}$ until reaching a peak value at $\sigma_{peak}$. This magnetization corresponds to the value at which the inversion contains the highest energy fraction for a given temperature. However, $f_{inv}$ then decreases as $\sigma_{rel}$ increases up to $\sigma_0$, which we define as the magnetization at which $I_2/I_1 = 1$. As discussed in Section \ref{sec:Alfven}, this value is obtained when $v_{A,rel} \rightarrow c$. The change in behaviour at lower temperatures can be explained by noting that at $\sigma_0$, the direction of $\left(\frac{\partial F_0}{\partial t}\right)_{\max}$ is almost exactly perpendicular for $\theta = 10^{-2}$, as is shown in Fig. \ref{fig:angle}. This is not the most efficient angle for the formation of the crescent shape, as particles need to be accelerated to larger values of $|q_\perp|$ in the region $q_\parallel >0$, rather than in the region centred at $q_\parallel = 0$. 

When the magnetization $\sigma_{rel} > \sigma_0$, the ratios of the various terms in equation \eqref{eq:4} are constant because $I_1$, $I_2$ and $I_3$ are unchanging. Therefore, the fraction of energy in the inversion is independent of $\sigma_{rel}$ when $\sigma_{rel} > \sigma_0$. This region is shown by the horizontal bars in Fig. \ref{fig:T3} to \ref{fig:Tm2}, which continue indefinitely to higher values of $\sigma_{rel}$. For temperatures of $\theta \geq 10^{-1}$, $\sigma_0$ also corresponds to $\sigma_{peak}$, as the ratio $\left(\frac{\partial F_0}{\partial t}\right)_{mix,\max}/\left(\frac{\partial F_0}{\partial t}\right)_{\perp,\max}$ is at its greatest for $I_2/I_1 = 1$. Therefore, for these temperatures $f_{inv}$ is at its maximum value for all $\sigma_{rel} \geq \sigma_{0}$.
	
Below $\sigma_1$, the plasma is in the intermediate magnetization region where the fraction of energy in the population inversion increases linearly with $\sigma_{rel}$. As in the high magnetization regime, this behaviour can be explained by the relative strength of the mixed term, as $\left(\frac{\partial F_0}{\partial t}\right)_{mix,\max}/\left(\frac{\partial F_0}{\partial t}\right)_{\perp,\max}$ is also proportional to $\sigma_{rel}$ for $\sigma_{rel}<\sigma_1$. This reduction in $f_{inv}$ corresponds to the weakening of the pitch angle diffusion as the magnetization decreases (and plasma $\beta$ increases), as discussed in Section \ref{sec:alpha}. In this regime the crescent shape no longer forms as clearly as in the high magnetization regime, as can be seen by comparing Fig. \ref{fig:dfalli10}, which shows $\frac{\partial F_0}{\partial t}$ in this intermediate magnetization regime, to Fig. \ref{fig:dfalli2} and \ref{fig:dfalli3} which have $\sigma_{rel} > \sigma_1$.

\begin{figure}
		\centering
		\includegraphics[width=\columnwidth]{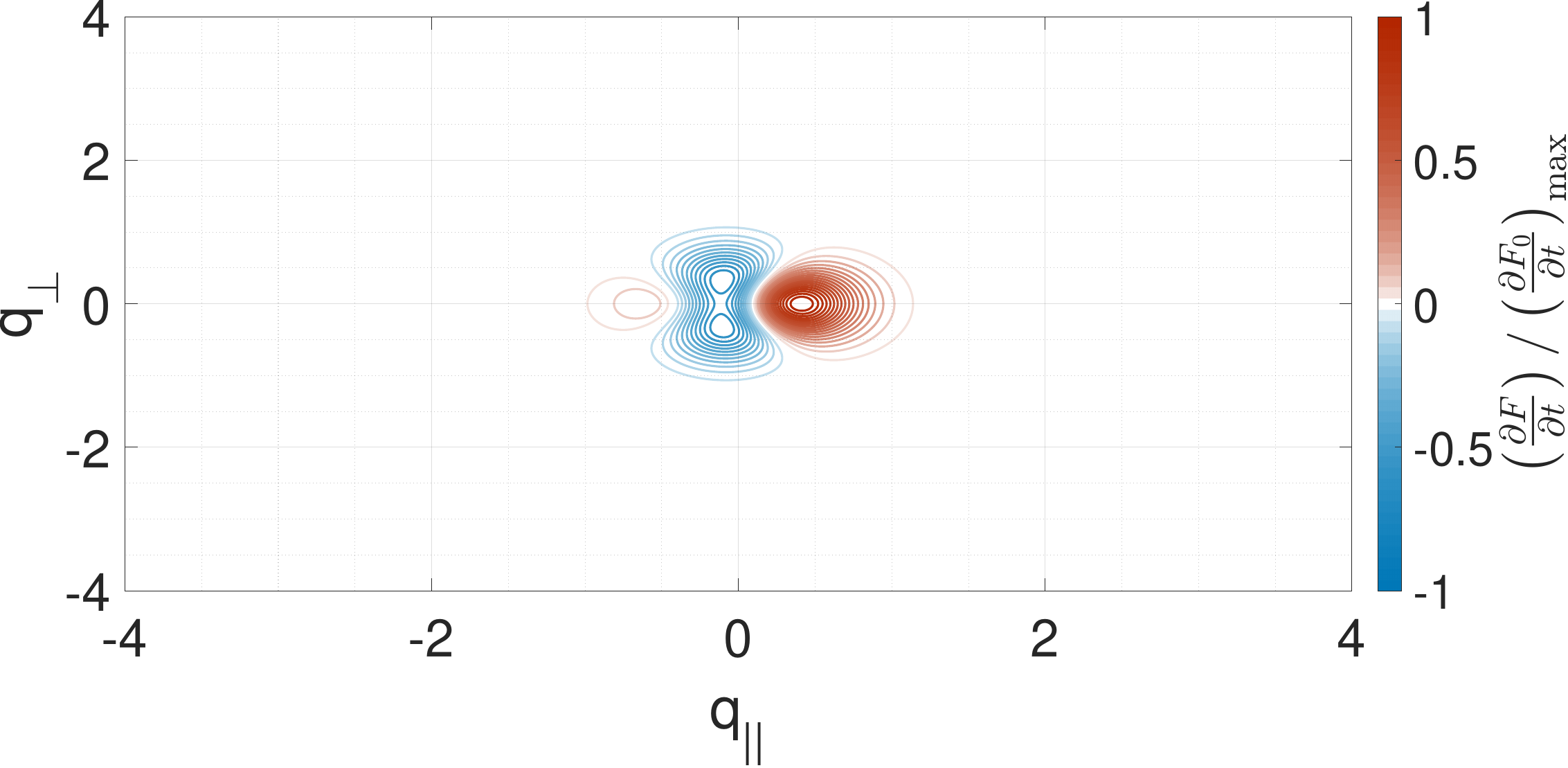}
		\caption{Contour plot of $\frac{\partial F}{\partial t}$ for an anisotropic Maxwell-J\"{u}ttner distribution in the low magnetization regime with $\theta = 0.1$, $I_2/I_1 = 30$ and $A=1.1$. Note the appearance of the positive region of $\frac{\partial F}{\partial t}$ centred at $q_\perp=0$ for $q_\parallel < 0$, which is completely absent when no temperature anisotropy is present.  }
		\label{fig:c4}
	\end{figure}

The $f_{inv} \propto \sigma_{rel}$ relationship holds for all $\sigma_1 > \sigma_{rel} > \sigma_2$. Here, $\sigma_2$ is the magnetization value below which a cut-off exists. 
This effect occurs due to the appearance of a positive region of $\frac{\partial F}{\partial t}$ centred at $q_\perp=0$ and $q_\parallel<0$ once the magnetization drops below this cut-off value. 
Above this threshold there is little to no diffusion in the parallel direction, with particles rather being moved to larger positive $q_\parallel$ values only. However, below $\sigma_2$, particles begin to spread in both the positive and negative $q_\parallel$ direction, sharply reducing the fraction of particles being added to the population inversion. The contrast in $\frac{\partial F}{\partial t}$ between the two scenarios is visible by comparing Fig. \ref{fig:dfalli30}, which presents $\frac{\partial F}{\partial t}$ for $A=1$, to Fig. \ref{fig:c4}, which shows $\frac{\partial F}{\partial t}$ for $A=1.1$. $I_2/I_1=30$ and $\theta = 10^{-1}$ in both cases. In order for this effect to occur, a temperature anisotropy with $\theta_\perp > \theta_\parallel$ is required, as is shown below. Such an anisotropy always occurs due to the nonresonant interaction \citep{2006PhRvL..96l5001W}, and the cut-off effect acts to reduce this anisotropy by increasing $\theta_\parallel$. As discussed in Section \ref{sec:alpha}, the strength of $A$ depends on both the magnetization and temperature. For the low magnetization regime in question here, the temperature anisotropy is generally small, with $1 \lesssim A < 2$.

 	 \begin{table}
	 	\centering
   	 	\caption{A summary of the relationship between $f_{inv}$ and $\sigma_{rel}$ in the different magnetization regimes}
	 	\label{tab:regimes}
	 	\begin{tabular}{ll}
	 		\hline
	 		magnetization & $f_{inv}$ \\ \hline
	 		$ \sigma_0 < \sigma_{rel} $                   & Constant    \\
	 		$\sigma_1 < \sigma_{rel} < \sigma_0$                    & Weakly dependent on $\sigma_{rel}$      \\
	 		$\sigma_2 < \sigma_{rel} < \sigma_1$                  & $f_{inv} \propto\sigma_{rel}$    \\
	 		$\sigma_{rel} < \sigma_2$                 & Decreases sharply as $\sigma_{rel}$ decreases    
	 	\end{tabular}

	 \end{table}

\begin{figure}
	 	\centering
	 	\includegraphics[width=\columnwidth]{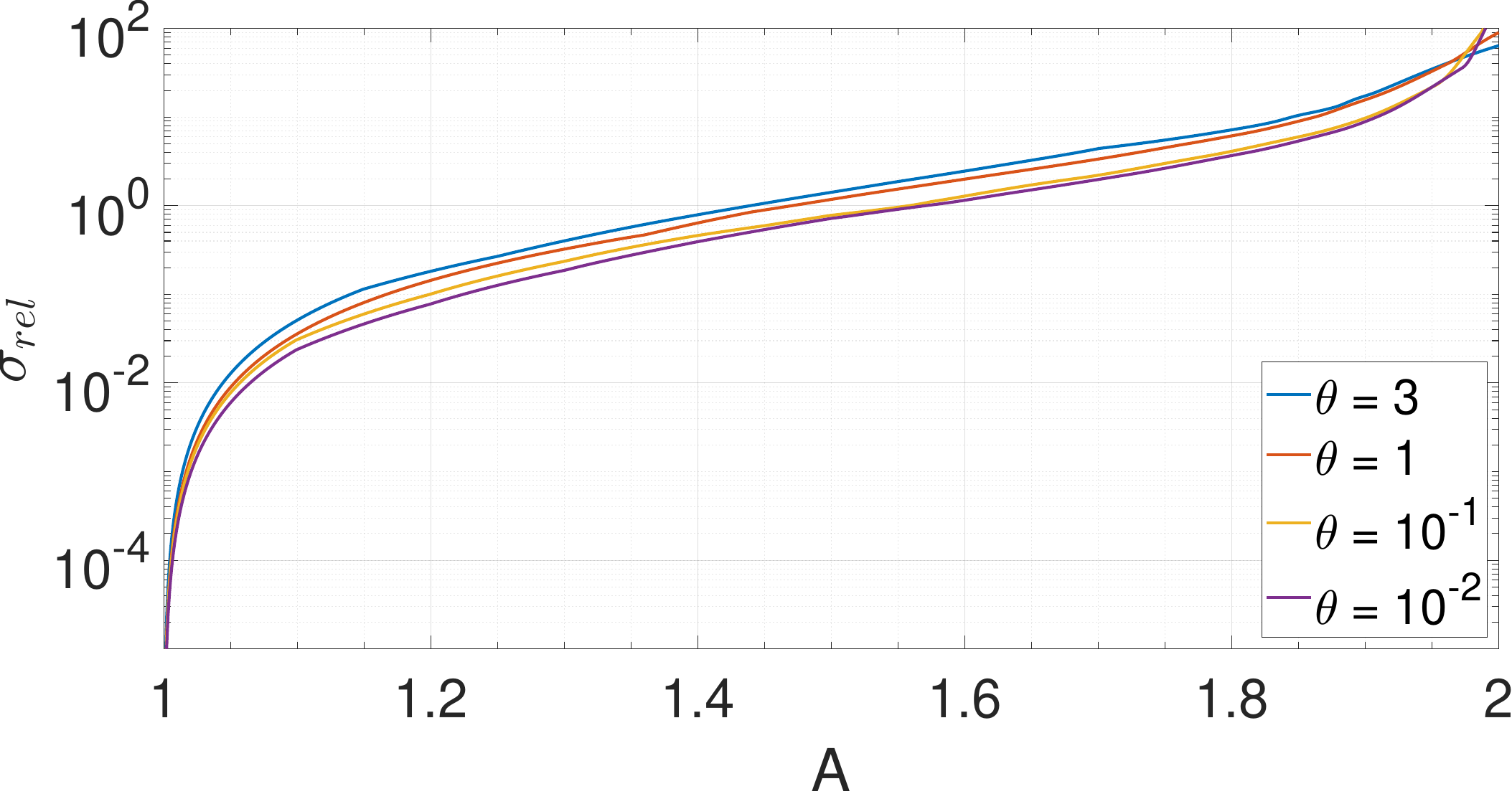}
	 	\caption{The maximum value of $\sigma_{rel}$ for which a positive $\frac{\partial F}{\partial t}$ in the $q_\parallel<0$ region will appear, plotted as a function of the temperature anisotropy $A$. A value of $\sigma_{rel}\sim10^2$ corresponds to $I_2/I_1=1$. It can be seen that as the magnetization increases, the required anisotropy $A$ for the cut-off effect to occur also increases from a value of $A\sim 1$ for $\sigma_{rel}\lesssim 10^{-4}$ to $A\sim2$ for $\sigma_{rel}\sim 10^2$. The relationship depends only weakly on the temperature, with the effect becoming relevant for very small temperature anisotropies of close to 1 at low magnetizations in all cases. }
	 	\label{fig:sigma_cut-off}
	 \end{figure}

	 An approximation for the value of $\sigma_2$ can be found by determining when $\frac{\partial F}{\partial t}$ first becomes positive for $q_\perp=0$ and $q_\parallel<0$. For the purpose of the following analysis, we assume an anisotropic Maxwell-J\"{u}ttner distribution of the form \citep{2023FrP....1174557B}
	 
	 \begin{equation}
	 F = C \exp\left(-\theta_\perp^{-1}\sqrt{1+q_\perp^2 + A q_\parallel^2}\right), \label{eq:amj}
	 \end{equation}
	where $C$ is the normalization factor. While this does not capture the full crescent shape expected at high magnetizations, it is a good approximation for low magnetizations where the fraction of particles in the inverted population is small.  The expression for $\frac{\partial F(q_\perp = 0)}{\partial t}$ is taken from Eq. \ref{eq:4}:
	 		\begin{align}
	 \frac{\partial F(q_\perp = 0)}{\partial t}&=-C\frac{\exp\left(-\theta_\perp^{-1}\sqrt{1+Aq_\parallel^2}\right)}{\theta_\perp(1+q_\parallel^2)^{3/2}\sqrt{1+A q_{\parallel}^2}} \nonumber \\
	 &\times	\Big(-(A-1)I_2q_\parallel^2\sqrt{1+q_\parallel^2} + (A-2)I_1q_\parallel(1+q_\parallel^2) \nonumber \\
  &+I_3(1+q_\parallel^2)^{3/2}\Big),
	 \end{align}
	 where $C$ contains all the constants omitted for clarity. Setting $\frac{\partial F(q_\perp = 0)}{\partial t} = 0$ results in the expression
	 \begin{equation}
	 -(A-1)I_2q_\parallel^2\sqrt{1+q_\parallel^2} + (A-2)I_1q_\parallel(1+q_\parallel^2) +I_3(1+q_\parallel^2)^{3/2}=0.
	 \end{equation}
	 The solution for $q_\parallel<0$ in the linear regime where $I_1 = \frac{v_{A,rel}}{c}I_2 =\frac{c}{v_{A,rel}}I_3$  is then given by
	 \begin{equation}
	 q_\parallel = -\frac{1}{\sqrt{(I_2/I_1)^2(A-1)^2-1}}.
	 \end{equation}
	 A positive $\frac{\partial F}{\partial t}$ in the $q_\parallel<0$ region is only possible when $I_2/I_1>1/|A-1|$. As $I_2/I_1$ increases, the required anisotropy becomes smaller. In other words, this effect occurs sooner for lower magnetizations, resulting in the lower values of $f_{inv}$ as $\sigma_{rel}$ decreases. This relationship is only weakly dependent on the temperature, with the only effect occurring as a result of the temperature dependence of the $\sigma -I_2/I_1$ relationship, as shown in  Fig. \ref{fig:Isigma}. The maximum value of $\sigma_{rel}$ for which a positive $\frac{\partial F}{\partial t}$ in the $q_\parallel<0$ region will appear is plotted in Fig. \ref{fig:sigma_cut-off} as a function of the temperature anisotropy. The figure extends only to $\sigma_{rel} = 10^2$ as this is the approximate value at which $I_2/I_1 \rightarrow 1$. At low magnetizations the cut-off effect occurs for any anisotropy $A>1$. The effect will therefore have an impact on the evolution of the distribution almost immediately. The needed anisotropy increases as the magnetization increases, with $A=2$ required for $\sigma_{rel}>\sigma_0$ (equivalent to $I_2/I_1 =1$). This result means that the   positive $\frac{\partial F}{\partial t}$ in the $q_\parallel<0$ region will appear for all parameters provided a sufficient temperature anisotropy exists, though  in the high magnetization cases the crescent shape will have already largely formed before this occurs. The weak temperature dependence is also apparent in Fig. \ref{fig:sigma_cut-off}, with only a factor of $\sim2$ in difference in the values of $\sigma_{rel}$ for $\theta = 10^{-2}$ and $\theta = 3$. This result can also be seen in our numerical results, where $\sigma_2$ only varies weakly between temperatures. However, we note that the values of $\sigma_2$ shown in Fig. \ref{fig:T3} to \ref{fig:Tm2} are only approximations due to the transition region not being sharp.
	 
\begin{figure}
	 	\centering
	 	\includegraphics[width=\columnwidth]{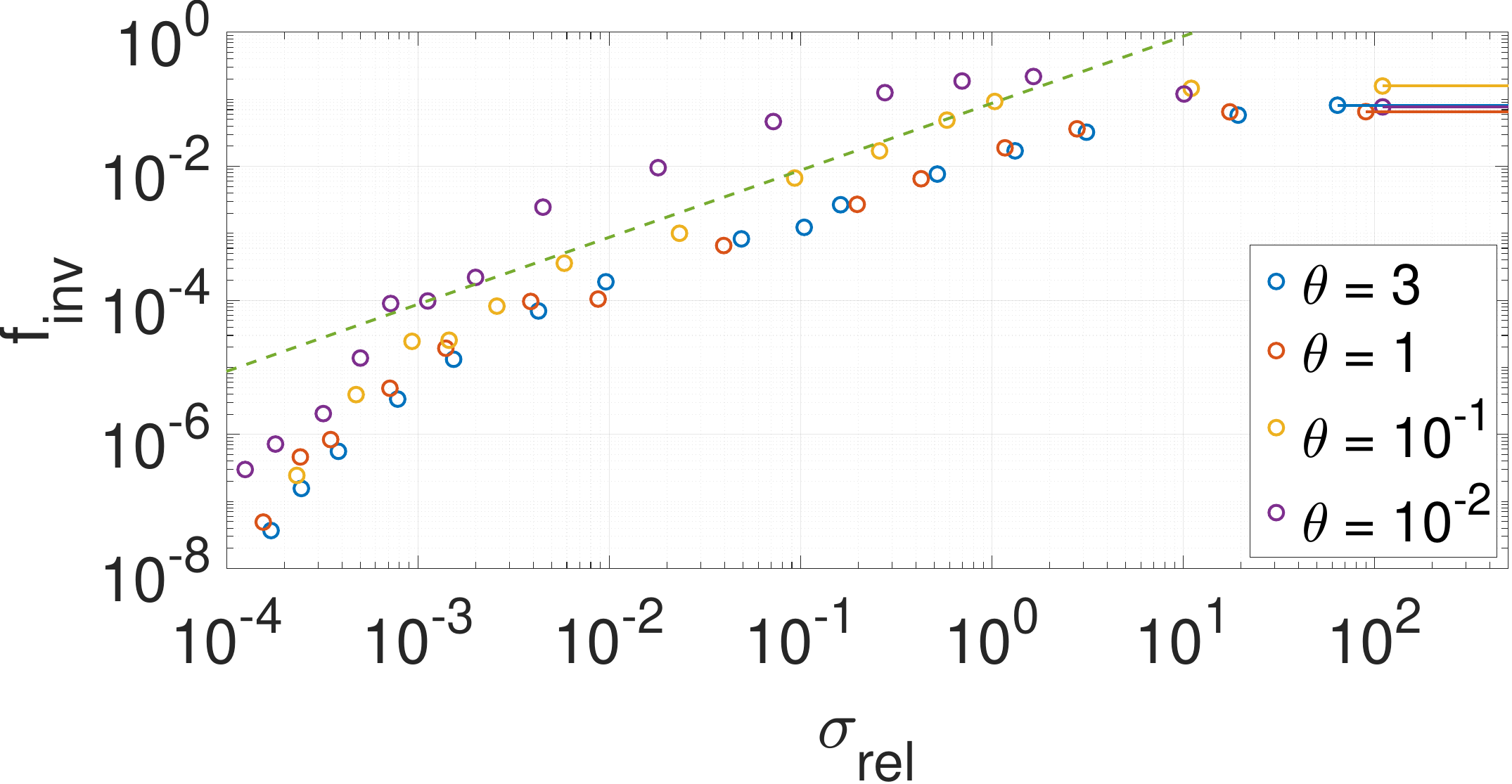}
	 	\caption{The energy in the inversion at $\eta =1$ versus magnetization for $\theta = 10^{-2}$ to $\theta=3$. The green dashed line shows a fit of $f_{inv} \propto \sigma_{rel}$ in the intermediate magnetization regime for $\theta = 1$. The value of $f_{inv}$ generally decreases as the temperature increases to relativistic values.}
	 	\label{fig:multisig}
	 \end{figure}

	 To examine the temperature dependence of the interaction, we plot all the results from Fig. \ref{fig:T3} to \ref{fig:Tm2} together in Fig. \ref{fig:multisig}, which shows $f_{inv}$ at $\eta = 1$ for every parameter set examined. There is a clear increase in $f_{inv}$ as the temperature decreases from relativistic values to $\theta = 10^{-2}$. The  dependence is most pronounced in the intermediate magnetization, as can be seen by examining the region $\sigma_{rel} \lesssim 1$ in Fig. \ref{fig:multisig}. For a given $\sigma_{rel}$, $f_{inv}$ increases as the temperature decreases, with a typical difference of over an order of magnitude between results for $\theta = 3$ and $\theta = 10^{-2}$. This indicates that the mechanism becomes less efficient in the relativistic regime at low magnetizations. As discussed in Section \ref{sec:alpha}, this is due to the decrease in relative strength of the pitch angle diffusion at higher temperatures. The weaker contribution from the mixed term at higher temperatures can be seen in Fig. \ref{fig:f4f5} and \ref{fig:f4f5sig}, as the relative strength of the mixed term for a given magnetization decreases as the temperature increases.

In the high magnetization regime, the difference in $f_{inv}$ for different temperatures is less significant. This is due to the fact that the strength of the mixed term reaches its maximum value at $\sigma_{rel}\sim 1$ rather than $\sigma_0$ for lower temperatures, as shown in Fig. \ref{fig:f4f5sig}. The precise values of $f_{inv}$ cannot be compared directly to the values of $\left(\frac{\partial F_0}{\partial t}\right)_{mix,\max}/\left(\frac{\partial F_0}{\partial t}\right)_{\perp,\max}$ as a turbulence level of $\eta>1$ is required to obtain the maximum population inversion in the majority of cases. However, in all cases values of $f_{inv}>10^{-2}$ are achieved in the high magnetization regime for both $\eta = 0.3$ and $\eta = 1$, showing that the nonresonant interaction is capable of producing a substantial population inversion across a wide range of both relativistic and non-relativistic temperatures. 

The turbulence level required to reach a given $f_{inv}$ is also temperature dependent, with lower temperatures requiring lower turbulence levels, as can be seen by the smaller differences between the values of $f_{inv}$ for $\eta = 1$ and $\eta=0.3$ at lower temperatures in Fig. \ref{fig:energies}. At higher temperatures the particles have more kinetic energy, and thus higher turbulence levels are needed to efficiently diffuse them. At lower magnetizations, the interaction is more sensitive to the turbulence level of the Alfv\'en waves, as can be seen by the $\eta=0.3$ values dropping more rapidly than those for $\eta =1$. In order for any significant distortion of the population distribution from a Bi-maxwellian in this part of the parameter space turbulence levels close to $\eta = 1$ are needed due to the inefficiency of pitch angle scattering in the low magnetization regime, as discussed above. 

The combination of the general decrease in $f_{inv}$ for lower magnetizations as well as the higher turbulence levels required to reach even these lower levels leads us to conclude that the parameter space $\sigma_{rel} > \sigma_1$ is the one of most relevance to SME, as here $f_{inv} \gtrsim 10^{-2}$ for all temperatures and a clear crescent shape is always formed as $\alpha > 0$.
	 
 \subsection{Time-scales} 
\label{sec:timescales}
The time taken for the population inversion to form is another important result as the mechanism must operate on a time-scale shorter than the characteristic dynamical time-scale to provide a valid explanation for FRBs. For an exponentially growing $\eta$, the time to reach a particular turbulence level is $t(\eta) = \left(2\Gamma\right)^{-1}\log(\eta/\eta_0)$. The time-scale therefore depends linearly on the growth rate. Due to the weak dependence on $\eta/\eta_0$, this time-scale is always on the order of $t\Gamma\sim \text{a few seconds}$. We note that as the turbulence approaches its maximum value, $\frac{\partial\eta}{\partial t}$ will naturally decrease, increasing the time taken for the final part of the evolution. Assuming that the Alfv\'en wave turbulence emerges from the central magnetar rather than the plasma in the emission region, the value of $\frac{\partial \eta}{\partial t}$ is independent of the plasma temperature. As a result the time-scale only depends on temperature through the lower turbulence levels required at lower temperatures. This therefore results in only a minor difference in time-scales between the results for $\theta = 3$ and $\theta = 10^{-2}$. While the turbulence level is independent of the Alfv\'en wave spectrum, the time-scale will be affected by the value of $\Gamma_k$ and whether $\Gamma_k$ is constant or varies with wave-number. The time-scale for a specific spectrum can be retrieved using the details described in Section \ref{sec:spectrum}. To explore the validity of these time-scales in the context of FRBs we compare the time-scales described here to those expected from magnetars in Section \ref{sec:mtime} below.

\section{Discussion}
\label{sec:discussion}
\subsection{SME efficiency and other plasma instabilities}
The numerical results presented in Section \ref{sec:num} above demonstrate that energy fractions of up to $f_{inv}\gtrsim 10^{-2}$ are attainable at high magnetizations through the nonresonant interaction between Alfv\'en waves and relativistic plasma. The energy fraction decreases linearly with $\sigma_{rel}$ in the intermediate regime $\sigma_2 < \sigma_{rel} < \sigma_1$ before dropping sharply below the cut-off $\sigma_2$. The time-scale in all cases is $t\sim  (2\Gamma)^{-1}\log(\eta/\eta_0)10^{-9}\text{ s}$. The above results are dependent on the temperature, with approximately an order of magnitude difference in $f_{inv}$ between $\theta = 10^{-2}$ and $\theta = 3$ in the low and intermediate magnetization regimes for $\eta = 1$. The energy in the inversion depends strongly on the energy density of the Alfv\'en waves, with values of $\eta\gtrsim0.1$ required to produce significant effects. We discuss the applicability of these parameters to magnetars and other neutron stars in Section \ref{sec:physical}, as these are the leading candidates for FRB progenitors, with one FRB directly observed to have originated from the galactic magnetar SGR 1935+2154 \citep{2020Natur.587...54C}.

 The results for $f_{inv}$ summarized above and presented in detail in Section \ref{sec:energy} can be contrasted with the efficiency results obtained from PIC simulations of relativistic shocks, another important candidate in explaining SME in relativistic environments \citep[e.g.][]{1992ApJ...391...73G, 2006ApJ...653..325A, Plotnikov2019}. In the relativistic shock model, the population inversion is not formed through nonresonant interactions with Alfv\'en wave. Instead, a soliton-like structure forms at the front of a shock with high magnetization. In this structure, the particles gyrating around the enhanced magnetic field near the shock front form a semi-coherent ring-like distribution in momentum space which can support SME \cite[e.g.][]{1988PhFl...31..839A, 1992ApJ...391...73G, 2006ApJ...653..325A, Plotnikov2019}. 
  
Comparing the results obtained for $f_{inv}$ from the nonresonant interaction to the efficiency values from the relativistic shock model shows several differences in the viable parameter spaces. In the high magnetization regime, the results presented above in Section \ref{sec:num} produce considerably higher $f_{inv}$ than in the case of relativistic shocks, where an efficiency of $\sim10^{-3}\sigma_{rel}^{-1}$ for $\sigma_{rel}>>1$ was found \citep{Sironi2021}. On the other hand, in the intermediate magnetization regime, for a magnetization of $\sigma_{rel}\sim0.1$ the shock efficiency was found to be $\sim 0.1$ \citep{Plotnikov2019}, comparable to the values found in this work for $\theta > 10^{-2}$. These are approximately $f_{inv}\sim 2\times10^{-2}\sigma_{rel}$ for $\sigma_2<\sigma_{rel}<1$ for $\theta = 1$ and $\theta = 3$, increasing to $f_{inv}\sim 0.5\sigma_{rel}$ for $\sigma_2<\sigma_{rel}\lesssim0.3$ at the lower temperature of $\theta = 10^{-2}$. These values are obtained assuming $\eta = 1$, and decrease at lower values of $\eta$. These results suggest that nonresonant interactions with Alfv\'{e}n waves may be the more viable mechanism at high magnetizations, while relativistic shocks, if exist, may be more relevant at lower magnetizations.
  
We emphasize that the efficiency in the relativistic shock scenario and $f_{inv}$ are not the same quantity. While the efficiency values above describe the fraction of energy actually emitted by SME, $f_{inv}$ provides an upper limit on the total energy fraction available from the deformed distribution. Not all of this energy will necessarily be extracted due to the influence of other plasma processes. For instance, at higher temperatures the SME growth rate may be reduced \citep{2006ApJ...653..325A, 2017ApJ...840...52I, 2020MNRAS.499.2884B}.
However, this may not affect the overall efficiency, provided that all other plasma instabilities have a slower growth rate.  

  
The impact of plasma instabilities is most relevant for the high temperature and low magnetization regions of the parameter space. As noted above, the particle distribution can be roughly approximated as a Bi-Maxwellian, especially in the low magnetization regime where the fraction of energy in the inversion is small. In such distributions both the firehose and mirror instabilities are important. These have stability criteria  of $A > 1-1/\beta_\parallel$  for the firehose instability and $A < 1+1/\beta_\parallel$ for the mirror instability \citep{gary1993theory}. Here $\beta_\parallel = 2 \theta_\parallel/\sigma_{rel}$ is the parallel plasma beta. When $\sigma_{rel}>>1$, $\beta_\parallel<<1$ for all temperatures, resulting in a distribution which is not susceptible to either of the instabilities. However, when $\sigma_{rel} << 1$ and the temperature is relativistic, even a small temperature anisotropy will result in the stability criteria no longer being satisfied. In the case of the low magnetization regime, the temperature anisotropy is typically in the range $1\lesssim A \lesssim 2$, and so the distribution will be potentially affected by the mirror instability at high temperatures. This means that plasmas in the low and intermediate magnetization regimes have both lower energy fractions and are more susceptible to instabilities. On the other hand, this issue has no effect on the most efficient parameters, namely those in the high magnetization regime, strengthening the conclusion that this is the regime of most relevance for SME in astrophysical scenarios.

\subsection{Magnetar time-scales}
\label{sec:mtime}
As magnetars are the primary candidate for FRB progenitors  \citep{2020Natur.587...54C}, we consider the typical time-scales of such objects. Magnetars have two time-scales of interest, namely the radius and period of the star, either of which could potentially provide the crucial time variability in the Alfv\'{e}n wave field which is a requirement for a nonresonant interaction. Firstly, the magnetar radius $R_*$ corresponds to a time-scale of $\sim R_*/c = 3.3\times 10^{-5}R_{*,6}\text{ s}$. The time-scale of the magnetar rotational period is significantly longer, typically $\sim 1-10\text{ s}$ \citep{2008A&ARv..15..225M, 2014ApJS..212....6O}, \footnote{https://www.physics.mcgill.ca/~pulsar/magnetar/main.html},
though young magnetars may have millisecond periods \citep{2022ASSL..465..245D}.

Considering the Alfv\'en wave turbulence grows on a time-scale which is some fraction $\chi$ of these magnetar time-scales, the time to reach a turbulence value of $\eta$ is $t\sim 1.7\times10^{-5} \chi^{-1}R_{*,6}\log(\eta/\eta_0)\text{ s}$ for the magnetar radius time-scale. The period scale produces longer times of $t \sim 0.5 \chi^{-1}P\log(\eta/\eta_0)$. While the period time-scale is considerably longer, it naturally has a wider range of possible values due to the variation in neutron star periods. In both cases the constraint that $\Gamma << \Omega$ is comfortably satisfied in the FRB emission regions for a wide range of $\chi$. These time-scales are applicable across the whole temperature and magnetization parameter space, with only a weak dependence on temperature in the form of the increased turbulence levels required when the initial temperature is higher.

 The time-scale of the mechanism is especially important to FRBs as it is unclear if all sources repeat or not. While some studies show observational differences between apparent one-off bursts and repeaters \citep{2021ApJ...923....1P, 2022A&ARv..30....2P}, these one-off bursts may also be rarer, brighter events than other undetected signals from the same object \citep{2024NatAs...8..337K}. For potential one-off bursts, the time taken for the inversion to form is not a limiting factor. On the other hand, if most or all FRBs repeat, the inversion must form with times shorter then the typical time between bursts. While this model can not intrinsically produce time-scales as short as the $\sim 60\text{ ns}$ observed from the fastest repeaters \citep{2022NatAs...6..393N}, the vast majority of FRBs repeat on much longer time-scales, typically on the order of seconds or longer \citep{2022ApJ...927...59L, 2023ApJS..269...17H}. Variability on a very short time scale ($< \mu\text{s}$) may also be induced by the propagation of the FRB signal through a highly magnetized plasma, rather than as a result of the emission process itself \citep{2024arXiv240904127S}. These factors ensure that the mechanism is not limited to scenarios with extremely high values of $\Gamma$ as shorter time-scales would, suggesting that magnetars are viable sources for this mechanism to produce the majority of FRBs.

\subsection{Physical scenarios for FRBs}
\label{sec:physical}

As mentioned above, magnetars are the most likely progenitors of FRBs as they naturally provide the large energies and small size and time scales necessary to produce the high brightness temperatures observed from FRBs \citep[e.g.][]{Zhang2020}. FRB 200428 has also been observed to originate from the galactic magnetar SGR 1935+2154 \citep{2020Natur.587...54C}. In our model, we assume that the required Alfv\'en waves for the nonresonant interaction are produced by a quake or similar instability of the central magnetar \citep[e.g][]{1989ApJ...343..839B,2023ApJ...959...34B}. As the magnetic field of the Alfv\'en wave decreases more slowly than the magnetospheric field, the relative strength of the turbulence grows as $\eta = \eta_*(R/R_*)^{3/2}$ \citep[e.g.][]{2020ApJ...900L..21Y}. Here, $\eta_*$ is the turbulence level at the magnetar surface. To reach values of $\eta>0.1$ at the light cylinder therefore requires only small initial values of $\eta_*>3\times10^{-7}P^{-3/2}R_{*,6}^{3/2}$, suggesting that the turbulence levels needed to form a population inversion are achievable in this scenario.

 The Alfv\'en waves propagate outwards before interacting with a relativistic plasma, with the resulting nonresonant interaction producing a population inversion and SME. In order for this to produce an FRB signal, the interaction must occur in a region of parameter space where a high value of $f_{inv}$ can be achieved on a time-scale compatible with the dynamical time-scales discussed in Section \ref{sec:mtime}. The plasma parameters must also allow a peak SME frequency appropriate for FRBs, whose signals have primarily been detected in the GHz band with frequencies ranging from 110 MHz \citep{2019ARep...63...39F,2021ApJ...911L...3P,2021Natur.596..505P} to 8 GHz \citep{2018ApJ...863....2G}.

The plasma parameters required to achieve such SME frequencies depend on the magnetization of the plasma. At higher magnetizations of $\sigma_{rel}>1$, the  maser's peak frequency is $\omega_m\sim\Omega\gamma_{av}^{-1}$ \citep{Lyubarsky2021}. If the plasma where the emission occurs is moving towards the observer with Lorentz factor $\gamma_B$, this results in a required magnetic field of $B_0 \sim 360 \nu_9\gamma_{av}\gamma_B^{-1} \text{ G}$. Here $\nu$ is the FRB frequency in Hz. In order to satisfy $\sigma_{rel}>1$, the upper limit on the number density in this regime is $n_{\max}\sim1.2\times10^{10}\nu_9^2\gamma_{av}\gamma_B^{-2}\text{ cm}^{-3}$. On the other hand, when $\sigma_{rel}<1$, the peak frequency for SME occurs at  $\omega_m\approx\gamma_{av}^{-1/2}\omega_p \text{min}\left\lbrace\gamma_{av},\sigma_{rel}^{-1/4}\right\rbrace$ \citep{2002ApJ...574..861S}. As $\omega_m$ is only very weakly dependent on the magnetization, the peak frequency will never be much higher than a few times $\gamma_{av}^{-1/2}\omega_p$. The number density required to produce emission at FRB frequencies is therefore $n\sim1.2\times10^{10}\nu_9^2\gamma_{av}\gamma_B^{-2}\text{max}\left\lbrace\gamma_{av}^{-2},\sigma_{rel}^{1/2}\right\rbrace\text{ cm}^{-3}$, with a corresponding upper limit on the magnetic field of $B_{0,\max} \sim360 \nu_9\gamma_{av}\gamma_B^{-1}\text{max}\left\lbrace\gamma_{av}^{-1},\sigma_{rel}^{1/4}\right\rbrace\text{ G}$ to satisfy $\sigma_{rel}<1$. As the values for the magnetic field in the $\sigma_{rel} < 1$ case and the number density in the $\sigma_{rel}>1$ case are both upper limits, a large parameter space is viable for SME.

As the high magnetization regime produces the largest values of $f_{inv}$ and is less susceptible to plasma instabilities we first examine a model in this regime by considering Alfv\'en waves propagating through a relativistic magnetar wind. These winds have typical magnetizations of $  \sigma_{wind}\sim280B_{*,15}^{8/9}R_{*,6}^{24/9}M_3^{-2/3}P^{-2}
$ outside the light cylinder radius $R_{LC} = cP/2\pi = 4.8\times10^9P\text{ cm}$ \citep{Lyubarsky2021}. Here $B_*$ is the surface magnetic field and $M$ is the pair multiplicity. This magnetization value is well above $\sigma_1$ for all temperatures, and so is in the high magnetization regime with a correspondingly high $f_{inv}\gtrsim 10^{-2}$ at all temperatures examined. Furthermore as $f_{inv}$ is close to constant for all $\sigma_{rel}$ in this regime, the precise magnetization value of the wind does not effect the validity of the mechanism provided that $\sigma_{wind}>\sigma_{1}$. The distance from the neutron star at which emission could occur in the FRB frequency range is $R_{FRB}\sim1.2\times10^{11}\nu_9^{-1}\gamma_B\gamma_{av}^{-1}B_{*,15}R_{*,6}^3P^{-2}\text{ cm}$, assuming $B\propto1/R$ outside the light cylinder. This means that maser emission in this scenario occurs just outside the magnetosphere at radii approximately an order of magnitude smaller than in relativistic shock models, depending on the precise parameters of the FRB in question \citep{2020MNRAS.494.4627M}. As discussed above in Section \ref{sec:mtime} the high magnetization regime is restrictive in terms of the magnetar radius time-scale, especially at relativistic temperatures. This suggests that a longer scale such as the magnetar period is responsible for the variation in the Alfv\'{e}n waves in this model.  The  particle number density at this distance is significantly below the upper limits for $n_{\max}$ discussed above, so SME at the appropriate frequencies is viable in this scenario assuming the population inversion has successfully formed. 

This model also implies that SME may occur at smaller radii and thus higher frequencies than those observed from FRBs, as the background magnetic field is larger closer to the magnetar. Inside the magnetosphere at $R<R_{LC}$, $\sigma_{rel}>>1$ and so a population inversion with a large energy fraction could possibly be formed. However in this region $\eta$ will be much smaller than at larger radii outside the light cylinder due to the extremely high $B_0= 10^{15}B_{*,15} R_{*,6}^3/R_6^3 \text{ G}$ inside the magnetosphere. These low turbulence levels will not be sufficient to produce any significant change in the particle distribution through this mechanism.

Furthermore, strong radio emission may be damped due to non-linear effects at radii similar to and smaller than those required in the relativistic magnetized wind scenario \citep{2024arXiv240910732S}. This effect is important at $R\lesssim 2 \times 10^{12} L_{42}^{1/4}B_{*,15}^{1/2}R_{*,6}^{3/2}\nu_9^{-1}P^{-1}\text{ cm}$, where $L$ is the FRB isotropic luminosity (which has been observed to vary from $L\sim10^{38} \text{ erg s}^{-1}$ \citep{2020Natur.587...59B} to $L\sim10^{46} \text{ erg s}^{-1}$ \citep{2019Natur.572..352R,2023Sci...382..294R}). This constraint therefore requires $\gamma_B>>1$ to explain higher luminosity FRBs in the relativistic wind model. We note that the typical wind Lorentz factor is expected to be on the order of $\sim 90$ \citep{Lyubarsky2020,Lyubarsky2021}, satisfying this requirement. This constraint may also be weakened by the interaction of a flare from the original magnetar quake with the relativistic wind \citep[e.g.][]{2020ApJ...897....1L}. SME at much lower frequencies than FRBs is also not feasible in this model as the Alfv\'en wave frequency becomes comparable to $\Omega$ at larger radii, violating the conditions required for the nonresonant interaction to be dominant as described in Section \ref{sec:nonresonant}. The combination of these constraints leads us to conclude that nonresonant interaction between Alfv\'en waves and a relativistic magnetar wind leads to the appropriate conditions to produce FRBs at $R_{FRB}$. These limits may also provide information on the magnetar properties (such as period and magnetic field) based on the frequency and luminosity of observed FRB signals.

\subsection{Alternative FRB scenarios}
Magnetars may also be surrounded with a lower magnetization subrelativistic wind with $\beta_{wind}\lesssim 1$ and $\sigma_{wind}\lesssim 1$ \citep{Metzger2019}. Here $\beta_{wind} = v_{wind}/c$ is the normalized wind velocity. In this scenario wind densities of $   n \sim 1.6\times10^{5}\dot{\mathcal{M}}_{21}R_{14}^{-2}\beta_{wind}^{-1}\text{ cm}^{-3}$ are expected. Here, $\dot{\mathcal{M}}$ is the mass loss rate, which is normalized to constraints obtained from FRB 121102, where $\dot{\mathcal{M}}\sim10^{19}-10^{21}\text{ g s}^{-1}$ \citep{2018ApJ...868L...4M}. This environment satisfies the constraints for maser emission at a similar radius as the relativistic magnetar wind, $R_{FRB}\sim 3.7\times 10^{11}\nu_9^{-1}\dot{\mathcal{M}}_{21}^{1/2} \beta_{wind}^{-1/2} \gamma_B \gamma_{av}^{-1/2}\text{min}\left\lbrace\gamma_{av},\sigma_{rel}^{-1/4}\right\rbrace\text{ cm}$. This scenario is therefore also susceptible to the damping effect described above for high luminosity FRBs, as well as having considerably lower values of $f_{inv}$ than in the high magnetization case, with $f_{inv} \sim 0.1\sigma_{rel}$ before dropping precipitously for $\sigma_{rel} < \sigma_2$. 

Observations of rapidly varying polarization angles similar to pulsar emission have led to arguments in favour of a magnetospheric origin for FRBs \citep{2024arXiv240209304M, 2024ApJ...972L..20N}. SME typically produces emission with a high degree of linear polarization due to the ordered magnetic field lines in the emission region \citep[e.g.][]{ Plotnikov2019}, and does not intrinsically result in changes in polarization angle. However, recent works on SME from relativistic shocks suggest rapid swings of polarization angle are possible even far from a neutron star \citep{2024PhRvL.132c5201I}, and that proximity to the central engine is therefore not a requirement to match observations. As the polarization properties of FRBs are also subject to propagation effects this is an area which requires further examination.

Recent works have also suggested that magnetospheric models at shorter distances from the central engine are favoured over SME from longer distances due to limits obtained from scintillation measurements \citep{2024arXiv240611053N}. In both physical scenarios presented above, the distances $R_{FRB}$ are comparable to the upper limits of $R\sim 3\times 10^{11}\text{ cm}$ obtained from scintillation measurements of FRB 20221022A \citep{2024arXiv240611053N}. SME is thus not ruled out as an emission mechanism in these scenarios, though the limit is constraining on the parameters of the central object and wind. Emission at larger radii as in the models presented in this work also avoids potential issues with the propagation of the FRB signal out of the magnetosphere, as radio waves may be strongly damped and struggle to escape \citep[e.g.][]{2021ApJ...922L...7B, 2023ApJ...957..102G,2023arXiv230712182B}, though there is as of yet no consensus on this issue \citep[e.g.][]{2022MNRAS.515.2020Q,2024MNRAS.529.2180L}.

\section{Conclusion}
\label{sec:conclusion}
In this work, we showed that nonresonant interaction of Alfv\'{e}n waves with a relativistic plasma produce population inversions for temperatures in the range $10^{-2}\leq\theta\leq3$ and for magnetizations greater than $\sigma_{rel}\sim 10^{-4}$. We report that a significant fraction of the energy in the particle distribution is contained in the crescent shaped population inversion, with peak values of $f_{inv}\sim0.1$ reached at magnetizations of $\sigma_{rel}\gtrsim10$ for temperatures of $\theta \geq 10^{-1}$. At the lower temperature of $\theta = 10^{-2}$, the peak magnetization is lower at a value of $\sigma_{rel}\sim 0.1 - 1$, but similar energy fractions are obtained. In the intermediate magnetization regime with lower magnetizations of $\sigma_{rel}<\sigma_1\sim 10^{-1}$ the energy fraction decreases as the magnetization decreases, with $f_{inv}\propto \sigma_{rel}$, before dropping sharply at a cut-off magnetization of $\sigma_{rel}=\sigma_2\sim 10^{-4}$. The higher fractions of energy in the population inversion at high magnetizations across all temperatures, as well as reduced susceptibility to other instabilities, leads us to suggest that this regime is the most relevant for astrophysical scenarios such as FRBs. 

Significant turbulence levels of $\eta \gtrsim0.1$ are required to sufficiently distort the initial distribution, with the value of $\eta$ needed increasing with temperature and at reduced magnetizations.
The time-scale for the formation of the population inversion depends linearly on $\Gamma^{-1}$, but only weakly on the final turbulence level. In all cases the time-scale is therefore approximately $t \sim \Gamma^{-1}\text{ s}$. These values are compatible with the magnetar period and the light-crossing time-scale for all parameters. We have also demonstrated that the parameters at which the population inversion can form through nonresonant interactions are achievable in the environment of magnetars, and can lead to SME at the appropriate frequencies for FRBs. The FRB signal originates in the region close to but outside of the light cylinder, at distances of $R_{FRB}\sim1.2\times10^{11}\nu_9^{-1}\gamma_B\gamma_{av}^{-1}B_{*,15}R_{*,6}^3P^{-2}\text{ cm}$, and is able to escape the magnetar environment without significant damping. 

\section*{Acknowledgements}

AP acknowledges support from the European Union (EU) via ERC consolidator grant 773062 (O.M.J.). KL acknowledges the support of the Irish Research Council through grant number GOIPG/2017/1146 as well as funding obtained by the above ERC grant. The authors also  wish to acknowledge the Irish Centre for High-End Computing (ICHEC) for the provision of computational facilities and support. We would also like to thank Antoine Bret for helpful discussions regarding plasma instabilities, and the referee for their helpful comments.

\section*{Data Availability}
The code used to perform the calculations presented in this paper is available upon request.

\bibliographystyle{mnras}
\bibliography{paper_bibliography}

\appendix

\section{Semi-analytical examination of the terms in the quasilinear kinetic equation}
\label{sec:termcomp}
In this section we retrieve some semi-analytical approximations for the ratios of the various terms in equation \eqref{eq:4} in the low and intermediate magnetization regimes, specifically for $\sigma_{rel}<\sigma_1$, equivalent to $I_2/I_1>>1$.
\subsection{Mixed term}
\begin{figure}
\centering
\includegraphics[width=\columnwidth]{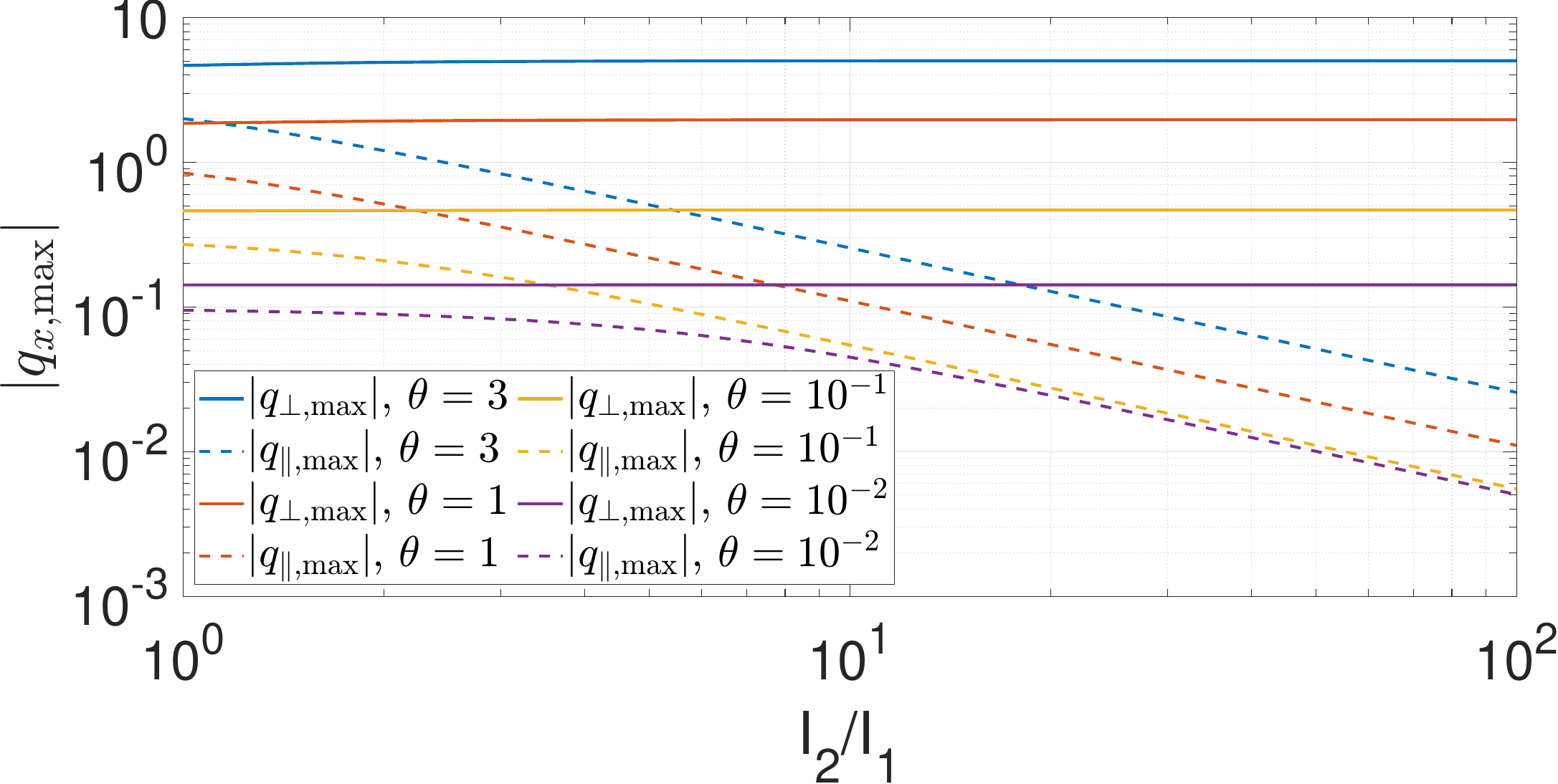}
\caption{The values of $|q_{\parallel,\max}|$ and $|q_{\perp,\max}|$ as a function of $I_2/I_1$ for the mixed term, $\left(\frac{\partial F}{\partial t}\right)_{mix}$. Dashed lines indicate $|q_{\parallel,\max}|$ at different temperatures while solid lines show $|q_{\perp,\max}|$.  For large $I_2/I_1>>1$, $|q_{\parallel,\max}|\propto (I_2/I_1)^{-1}$ and $|q_{\perp,\max}|$ is constant. }
\label{fig:maxmix}
\end{figure}

The change in the distribution due to the mixed term, $\left(\frac{\partial F_0}{\partial t}\right)_{mix}$, for an initial symmetric Maxwell-J\"{u}ttner distribution is given by
\begin{align}
\left(\frac{\partial F_0}{\partial t}\right)_{mix} &= \frac{C e^{-\sqrt{1+q_{||}^2+q_\perp^2}} q_{||}q_\perp^2}{\theta\left(1+q_{||}^2+q_\perp^2\right)^2}\left(1+\theta^{-1} \sqrt{1+q_{||}^2+q_\perp^2}\right) \nonumber \\
&\times\left(I_1\left(1+q_{||}^2+q_\perp^2\right)-I_2q_{||}\sqrt{1+q_{||}^2+q_\perp^2}\right),
\end{align}
where $C$ contains all the constants omitted for clarity. Fig. \ref{fig:maxmix} shows the values of $|q_{\parallel,\max}|$ and $|q_{\perp,\max}|$ as a function of $I_2/I_1$ for various temperatures, showing that $|q_{\perp,\max}|$ is close to constant and $|q_{||,\max}|\propto (I_2/I_1)^{-1}$ for large $I_2/I_1 >>1$ for all temperatures. It is also be seen that $|q_{||,\max}|<<1$ and $|q_{||,\max}|<<|q_{\perp,\max}|$ in this $I_2/I_1$ range. Therefore comparing $\left(\frac{\partial F}{\partial t}\right)_{mix,\max}$ at two different values of $(I_2/I_1)$ and $(I_2/I_1)'$ gives
\begin{align}
\frac{\left(\frac{\partial F}{\partial t}\right)_{mix,\max}}{\left(\frac{\partial F}{\partial t}\right)'_{mix,\max}}&\approx\left(\frac{q_{||,\max}}{q_{||,\max}'} \right)\nonumber \\
&\times\left(\frac{1+q_{\perp,\max}^2-(I_2/I_1)q_{||,\max}\sqrt{1+q_{\perp,\max}^2}}{1+q_{\perp,\max}^2-(I_2/I_1)'q_{||,\max}'\sqrt{1+q_{\perp,\max}^2}}\right).
\end{align}
Defining $C_\parallel$ as a constant of proportionality such that $q_{||,\max} = C_{\parallel}(I_2/I_1)^{-1}$, this expression can be simplified to

\begin{align}
\frac{\left(\frac{\partial F}{\partial t}\right)_{mix,\max}}{\left(\frac{\partial F}{\partial t}\right)'_{mix,\max}}&\approx\frac{C_{\parallel}(I_2/I_1)'\left(1+q_{\perp,\max}^2-C_{\parallel}\sqrt{1+q_{\perp,\max}^2}\right)}{C_{\parallel}(I_2/I_1)\left(1+q_{\perp,\max}^2-C_{\parallel}\sqrt{1+q_{\perp,\max}^2}\right)} \nonumber\\
&\approx\frac{(I_2/I_1)'}{(I_2/I_1)},
\end{align}
showing that $\left(\frac{\partial F}{\partial t}\right)_{mix,\max}\propto (I_2/I_1)^{-1}$ for large $I_2/I_1>>1$.

\subsection{Parallel terms}
\begin{figure}
\centering
\includegraphics[width=\columnwidth]{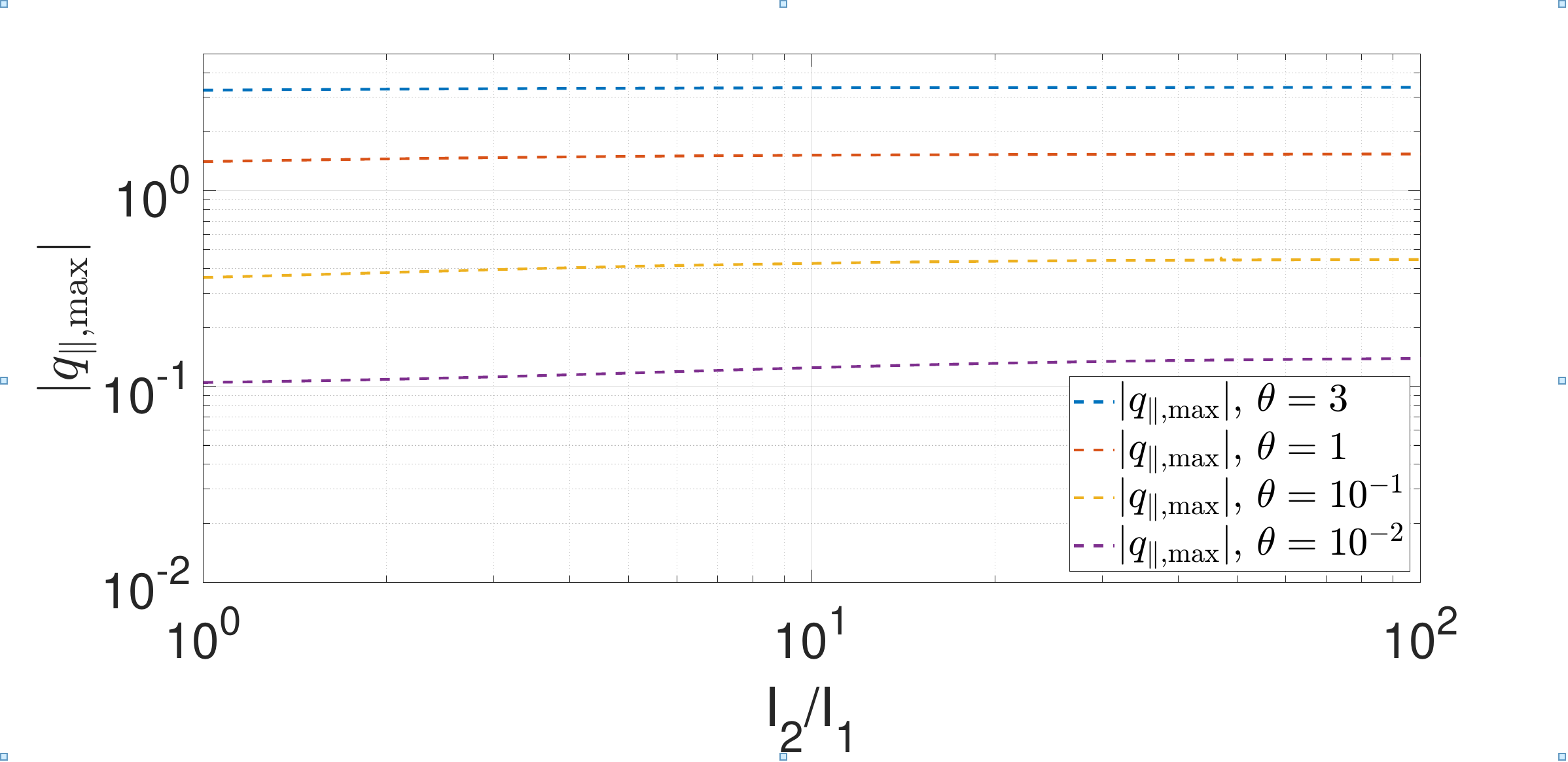}
\caption{The values of $|q_{\parallel,\max}|$ as a function of $I_2/I_1$ for the parallel advection term, $\left(\frac{\partial F}{\partial t}\right)_{\parallel,1}$. $|q_{\perp,\max}| = 0$ for all temperatures displayed.  For large $I_2/I_1>>1$, $|q_{\parallel,\max}|$ is approximately constant. }
\label{fig:maxpar}
\end{figure}

In the case of the parallel terms, the maximum value of the parallel advection term $\left(\frac{\partial F_0}{\partial t}\right)_{\parallel,1}$ is significantly higher than the diffusion term, as is shown in Fig. \ref{fig:termcomp}. For an initial symmetric Maxwell-J\"{u}ttner distribution, the term is
\begin{align}
\left(\frac{\partial F}{\partial t}\right)_{||,1} &= \frac{C e^{-\theta^{-1}\sqrt{1+q_{||}^2+q_\perp^2}}\theta^{-1} q_{||}}{\left(1+q_\parallel^2+q_\perp^2\right)} \nonumber \\
&\times\left(I_1(2+2q_\parallel^2+3q_\perp^2)-2 I_2 q_\parallel\sqrt{1+q_\parallel^2+q_\perp^2}\right),
\end{align}
where $C$ contains all the constants omitted for clarity. Fig. \ref{fig:maxpar} shows the values of $|q_{\parallel,\max}|$ as a function of $I_2/I_1$ for various temperatures, while $|q_{\perp,\max}| = 0$ for all cases. This allows us to discard all terms containing $q_{\perp,\max}$, resulting in
\begin{align}
\left(\frac{\partial F}{\partial t}\right)_{||,\max}&= \frac{C e^{-\theta^{-1}\sqrt{1+q_{||,\max}^2}}\theta^{-1} q_{||,\max}}{\left(1+q_{\parallel,\max}^2\right)} \nonumber \\
&\times\left(2 I_1(1+q_{\parallel,\max}^2)^2 -2 I_2q_{\parallel,\max}\sqrt{1+q_{\parallel,\max}^2}\right).
\end{align}
From Fig. \ref{fig:maxpar}, it is clear that $|q_{\parallel,\max}|$ is close to constant for a given temperature for large $I_2/I_1$. Therefore comparing $\left(\frac{\partial F}{\partial t}\right)_{||,\max}$ at two different values of $(I_2/I_1)$ and $(I_2/I_1)'$ gives

\begin{equation}
\frac{\left(\frac{\partial F}{\partial t}\right)_{||,\max}}{\left(\frac{\partial F}{\partial t}\right)'_{||,\max}}=\frac{(1+q_{\parallel,\max}^2)^2 -(I_2/I_1)q_{\parallel,\max}\sqrt{1+q_{\parallel,\max}^2}}{(1+q_{\parallel,\max}^2)^2-(I_2/I_1)'q_{\parallel,\max}\sqrt{1+q_{\parallel,\max}^2}}.
\end{equation}
 As the ratio  $\left|\frac{(1+q_{\parallel,\max}^2)^2}{q_{\parallel,\max}\sqrt{1+q_{\parallel,\max}^2}}\right| = \gamma(q_{\parallel,\max},0)^2/|\beta_{\parallel,\max}|< 15 $ for $\theta \leq 3$, the second term is always larger than the first for large values of $I_2/I_1$ and thus  $\left(\frac{\partial F}{\partial t}\right)_{||,\max}\propto I_2/I_1$ for $I_2/I_1>>1$.

\subsection{Perpendicular terms}

\begin{figure}
\centering
\includegraphics[width=\columnwidth]{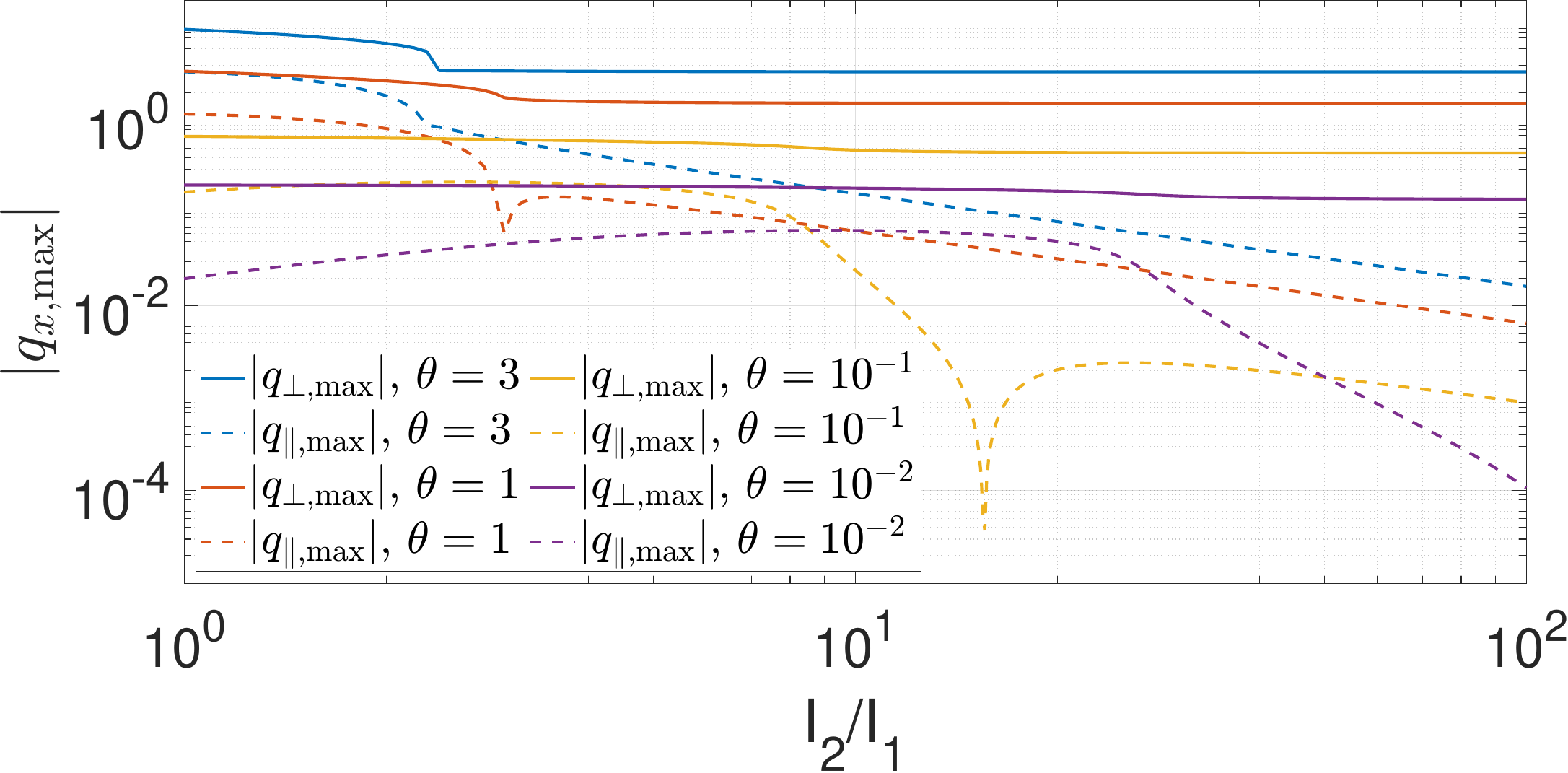}
\caption{The values of $|q_{\parallel,\max}|$ and $|q_{\perp,\max}|$ as a function of $I_2/I_1$ for the combined perpendicular terms $\left(\frac{\partial F}{\partial t}\right)_{\perp}$. Dashed lines indicate $|q_{\parallel,\max}|$ at different temperatures while solid lines show $|q_{\perp,\max}|$.  For large $I_2/I_1>>1$, $|q_{\parallel,\max}|<<1$ and $|q_{\perp,\max}|$ is approximately constant. The sharp dips in $|q_{\parallel,\max}|$ are due to a change in sign. }
\label{fig:maxperp}
\end{figure}

In the case of the perpendicular terms, both the advection and diffusion terms are significant, depending on the magnetization. For an initial symmetric Maxwell-J\"{u}ttner distribution, the sum of these terms is given by
\begin{align}
\left(\frac{\partial F}{\partial t}\right)_{\perp} &= \frac{Ce^{-\theta^{-1}\sqrt{1+q_{||}^2+q_\perp^2}}}{\theta^2\gamma^4}\Big\lbrace I_1 q_\parallel\gamma^2\left(-2q_\perp^2\gamma +4\theta(\gamma^2-q_\perp^2)\right) \nonumber \\
&+I_2[\theta \gamma q_\perp^2 (1+q_\perp^2) + q_\parallel^4(q_\perp^2 - 2\theta \gamma) \nonumber \\
&+q_\parallel^2(q_\perp^2 +q_\perp^4 - 2\theta \gamma)]\nonumber \\
&+I_3\gamma^2\left(q_\perp^4 - 2 \theta \gamma (1+q_\parallel^2) + q_\perp^2(1+q_\parallel^2 - 3\theta \gamma)\right)\Big\rbrace.
\end{align}
 Fig. \ref{fig:maxperp} shows the values of $|q_{\parallel,\max}|$ and $|q_{\perp,\max}|$ as a function of $I_2/I_1$ for various temperatures, showing that $|q_{\perp,\max}|$ is close to constant and $|q_{||,\max}|<<1$ and $|q_{||,\max}|<<|q_{\perp,\max}|$ for large $I_2/I_1$. We can therefore approximate $\left(\frac{\partial F}{\partial t}\right)_{\perp,\max}$ as 
\begin{align}
    \left(\frac{\partial F}{\partial t}\right)_{\perp,\max} &\approx \frac{Ce^{-\theta^{-1}\sqrt{1+q_{\perp,\max}^2}}}{\theta^2(1+q_{\perp,\max}^2)} \nonumber \\
    &\times\big\lbrace I_2 \theta q_{\perp,\max}^2 \nonumber \\
    &+I_3(q_{\perp,\max}^2\sqrt{1+q_{\perp,\max}^2}-\theta(2+3q_{\perp,\max}^2)\big\rbrace.
\end{align}
Now using the relation $I_3/I_1 = I_1/I_2$, which is true in the linear regime, comparing the value of $\left(\frac{\partial F}{\partial t}\right)_{\perp,\max}$ at two different values of $(I_2/I_1)'$ and $(I_2/I_1)$ gives

\begin{align}
& \frac{\left(\frac{\partial F}{\partial t}\right)_{\perp,\max}}{\left(\frac{\partial F}{\partial t}\right)'_{\perp,\max}}\approx\frac{(I_2/I_1)'}{(I_2/I_1)}\nonumber \\
 &\times\left(\frac{q_{\perp,\max}^2\sqrt{1+q_{\perp,\max}^2}+\theta(q_{\perp,\max}^2((I_2/I_1)^2-3)-2)}{q_{\perp,\max}^2\sqrt{1+q_{\perp,\max}^2}+\theta(q_{\perp,\max}^2((I_2/I_1)'^2-3)-2)} \right)\nonumber\\
&\approx\frac{(I_2/I_1)'}{(I_2/I_1)}\frac{q_{\perp,\max}^2\sqrt{1+q_{\perp,\max}^2}+\theta q_{\perp,\max}^2(I_2/I_1)^2}{q_{\perp,\max}^2\sqrt{1+q_{\perp,\max}^2}+\theta q_{\perp,\max}^2(I_2/I_1)'^2},
\end{align}
which is valid for $I_2/I_1>>1$. This ratio becomes 
\begin{equation}
     \frac{\left(\frac{\partial F}{\partial t}\right)_{\perp,\max}}{\left(\frac{\partial F}{\partial t}\right)'_{\perp,\max}} \approx \frac{(I_2/I_1)}{(I_2/I_1)'},
\end{equation}
when $\theta (I_2/I_1)^2>>\sqrt{1+q_{\perp,\max}^2} = \gamma(0,q_{\perp,\max})$. As $|q_{\perp,\max}|$ increases monotonically with temperature from $|q_{\perp,\max}|\sim0.14$ for $\theta = 10^{-2}$ to $|q_{\perp,\max}|\sim3.4$ for $\theta = 3$, this condition is satisfied in the $I_2/I_1>>1$ regime, confirming that $\left(\frac{\partial F}{\partial t}\right)_{\perp,\max}\propto I_2/I_1$ for low magnetizations.

\bsp	
\label{lastpage}
\end{document}